\documentstyle[12pt,aaspp4]{article}
\def\n1316{NGC~1316~}

\newcommand{\oiii}{[O\,{\sc III}]}

\newcommand{\etal}{{\it et al.}}
\newcommand{\PN}{{planetary nebulae~}}
\font\caps=cmcsc10			   
\newcommand{\rms}{{\caps RMS}}
\newcommand{\kms}{\mbox{km s$^{-1}$}}
\newcommand{\mpc}{\mbox{Mpc}}
\newcommand{\ppc}{\mbox{pc}}
\newcommand{\yr}{\mbox{years}}
\newcommand{\comment}[1]{}

\font\ninerm=cmr9  \font\sevenrm=cmr7  \font\sixrm=cmr5
  \font\fiverm=cmr5
\font\nineit=cmti9  
\font\ninesl=cmsl9
\font\ninei=cmmi9    \font\sixi=cmmi5
\skewchar\ninei='177
\font\ninesy=cmsy9  \font\sixsy=cmsy5
\skewchar\ninesy='60
\font\ninebf=cmbx9  \font\sixbf=cmbx5
\font\nineex=cmex10 scaled 833

\font\ninett=cmtt9
\def\adjustlinespace{\baselineskip=\baselineskip}
\def\ninepoint{\textfont0=\ninerm \scriptfont0=\sixrm 
                \def\rm{\fam0\ninerm}\relax
                \textfont1=\ninei \scriptfont1=\sixi 
                \def\mit{\fam1}\def\oldstyle{\fam1\ninei}\relax
                \textfont2=\ninesy \scriptfont2=\sixsy 
                \def\cal{\fam2}\relax
                \textfont3=\nineex \scriptfont3=\nineex 
                \def\it{\fam\itfam\nineit}\relax
                \textfont\itfam=\nineit
                \def\sl{\fam\slfam\ninesl}\relax
                \textfont\slfam=\ninesl
                \def\bf{\fam\bffam\ninebf}\relax
                \textfont\bffam=\ninebf \scriptfont\bffam=\sixbf
                \def\tt{\fam\ttfam\ninett}\relax
                \textfont\ttfam=\ninett
                \setbox\strutbox=\hbox{\vrule
                     hnine7pt depth2pt width0pt}\baselineskip=9pt
                \adjustlinespace
                \rm}
\font\caps=cmcsc10			   
\def\etal{{\it et~al.\ }}
\def\aa #1 #2 {A\&A, #1, #2}
\def\aas #1 #2 {A\&AS, #1, #2}
\def\acm #1 #2 {ACM-Trans Math Software, #1, #2}
\def\ada #1 #2 {Ann Astrophys, #1, #2}
\def\agabstr #1 #2 {Astr Ges Abstr Ser, #1, #2}
\def\aj #1 #2 {AJ, #1, #2}
\def\anach #1 #2 {Astr Nachr, #1, #2}
\def\apj #1 #2 {ApJ, #1, #2}
\def\apjl #1 #2 {ApJL, #1, #2}
\def\apjs #1 #2 {ApJS, #1, #2}
\def\araa #1 #2 {ARAA, #1, #2}
\def\apss #1 #2 {ApSpaceS, #1, #2}
\def\celmech #1 #2 {Cel Mech, #1, #2}
\def\esom #1 #2 {ESO Messenger, #1, #2}
\def\fundcp #1 #2 {FunCosP, #1, #2}
\def\jcp #1 #2 {J Comp Phys, #1, #2}
\def\jfm #1 #2 {J Fluid Mech, #1, #2}
\def\jmp #1 #2 {J Math Phys, #1, #2}
\def\ma #1 #2 {Mitt Astr Ges, #1, #2}
\def\mn #1 #2 {MNRAS, #1, #2}
\def\nat #1 #2 {Nat, #1, #2}
\def\obs #1 #2 {Observatory, #1, #2}
\def\pasj #1 #2 {PASJ, #1, #2}
\def\pasp #1 #2 {PASP, #1, #2}
\def\phyr #1 #2 {PhysRep, #1, #2}
\def\physd #1 #2 {Physica D, #1, #2}
\def\rpp #1 #2 {RepProgPhys, #1, #2}
\def\ssr #1 #2 {Sp Sci Rev, #1, #2}
\def\iau127#1{in de Zeeuw P.T. ed, Structure and Dynamics of 
     Elliptical Galaxies, IAU Symp.~No.~127. Reidel, Dordrecht, p.~#1}

\def\in#1#2#3#4#5#6{in: #1%
\if#2-%
\else%
, #2%
\fi%
\if#3-%
\else%
, ed.\ #3%
\fi%
\if#5-%
 {\if#4-%
 \else,%
   (#4)%
 \fi}%
\else%
 {\if#4-%
, (#5)%
\else%
, (#5:#4)%
\fi}%
\fi%
\if#6-%
.%
\else%
, #6.%
\fi%
}

\def\spose#1{\hbox to 0pt{#1\hss}}
\def\lta{\mathrel{\spose{\lower 3pt\hbox{$\mathchar"218$}}
     \raise 2.0pt\hbox{$\mathchar"13C$}}}
\def\gta{\mathrel{\spose{\lower 3pt\hbox{$\mathchar"218$}}
     \raise 2.0pt\hbox{$\mathchar"13E$}}}

\def\=#1{\overline{#1}}

\def\eps{\epsilon}

\def\pbyd#1#2{\mathchoice
             {\partial#1\over\partial#2}
             {\partial#1/\partial#2}
             {\partial#1\over\partial#2}
             {\partial#1\over\partial#2} }
\def\tbyd#1#2{\mathchoice
             {\hbox{d}#1\over\hbox{d}#2}
             {\hbox{d}#1/\hbox{d}#2}
             {\hbox{\sevenrm d}#1\over\hbox{\sevenrm d}#2}
             {\hbox{\fiverm d}#1\over\hbox{\fiverm d}#2} }

\def\kms{{\rm\,km\,s^{-1}}}

\def\kpc{{\rm\,kpc}}
\def\mpc{{\rm\,Mpc}}

\def\msun{{\rm\,M_\odot}}

\def\yr{{\rm\,yr}}

\def\rms{{\caps rms}}

\begin{document}

%
%

\title{The stellar dynamics and mass of \n1316\ using the radial velocities
of Planetary Nebulae\footnote{Based on observations 
made at the European Southern Observatory, La Silla, Chile, and
at the Siding Spring Observatory, NSW, Australia}}

\author{M. Arnaboldi }
\affil{Osservatorio Astronomico di Capodimonte, Naples 80131, Italy\\
magda@cerere.na.astro.it}

\author{K.C. Freeman}
\affil{Mt. Stromlo and Siding Spring Observatories, ACT 2611, Australia\\
kcf@mso.anu.edu.au}

\author{O. Gerhard, M. Matthias}
\affil{Astronomisches Institut, Universit\"at Basel, CH--4102 Binningen, 
Switzerland\\
gerhard@astro.unibas.ch,matthias@astro.unibas.ch}

\author{R.P. Kudritzki, R. H. Mendez}
\affil{Munich University Observatory, Munich 81679, Germany\\
kudritzki@usm.uni-muenchen.de, mendez@usm.uni-muenchen.de}

\author{M. Capaccioli}
\affil{Osservatorio Astronomico di Capodimonte, Napoli 80131, Italy\\
capaccioli@astrna.na.astro.it}

\author{H. Ford}
\affil{Physics and Astronomy Department, The John Hopkins University,
Baltimore 21218, U.S.A.\\
ford@jhufos.pha.jhu.edu}

\date{\today}


\begin{abstract}

We present a study of the kinematics of the outer regions of the
early-type galaxy NGC~1316, based on radial velocity measurements of
43 planetary nebulae as well as deep integrated-light absorption line spectra.

The smoothed velocity field of NGC 1316 indicates fast rotation at a
distance of 16 kpc, possibly associated with an elongated feature
orthogonal to the inner dust lanes. The mean square stellar velocity
is approximately independent of radius, and the estimated total mass
of the system is $ 2.9 \times 10^{11} M_\odot$ within a radius of 16 
kpc, implying an integrated mass--to--light ratio of M/L$_B \simeq 8$.

\end{abstract}


\section{Introduction}

The use of planetary nebulae (PNe) to measure radial velocities in the
outer regions of early-type galaxies is a powerful tool to extend
kinematical information beyond radii $r \simeq 2 R_e$, which are
presently out of reach for integrated light measurements (Carollo
\etal\ 1995, Gerhard \etal 1998). PNe are more suitable test particles
for this purpose than globular clusters (GCs) because the ionised
envelope of PNe can re-emit up to 15\% of the central star's energy in
the [OIII] $\lambda 5007$ line, the brightest optical emission line of
a PN (Dopita, Jacoby \& Vassiliadis 1992). With a high dispersion spectrum the
PN radial velocities are readily measured to an accuracy of 15 $\kms$.
Also, there are indications that the kinematics of globular clusters in
early-type galaxies may not be representative of the kinematics of the
underlying diffuse stellar population (eg. Hui \etal\ 1995, Grillmair \etal\
1994, Arnaboldi \etal\ 1994). While this need not affect the usefulness 
of globular clusters as dynamical tracers of the underlying mass, they may
give misleading impressions of the dynamical state of the outer stellar 
population in early-type galaxies; the angular momentum of the outer 
regions is particularly susceptible to mis-estimation in this way.

PNe have been used successfully as mass tracers in several nearby objects
and giant galaxies in the Virgo and Fornax clusters (see Arnaboldi \&
Freeman 1997 for a review). The best studied case, NGC 5128 (Hui
\etal\ 1995), for which 433 PN radial velocities were measured, shows
the power of the method: when the information from a sample of
discrete radial velocities is combined with the gas kinematics of the
inner regions, the viewing angles of the system and the mass
distribution can be estimated. Arnaboldi \etal\
1994, 1996 were able to measure tens of radial velocities with the ESO
New Technology Telescope (NTT) and Multi-Mode Instrument (EMMI)
spectrograph in the multi-object mode in the giant early-type galaxies
in the Virgo and Fornax clusters.  The size of these radial velocity
samples will be increased by an order of magnitude in the near future
with the advent of 8 meter telescopes and more efficient multi-object
spectrographs.

According to recent studies with penalised likelihood techniques
(Merritt \& Saha 1993, Merritt 1997), a few hundred to a thousand
radial velocities should be sufficient to constrain the gravitational
potential and the phase space distribution function in a
model-independent way. Much better results still can be expected if
large radial velocity samples are combined with deep integrated-light line
spectra, and the analysis is done simultaneously with line profile
modelling in the central $2R_e$ (Rix \etal 1997, Gerhard \etal 1998).

We present here a study of the galaxy NGC~1316, in
the Fornax cluster, also known as the radio continuum source
Fornax~A. The similarity between NGC~5128 (Cen~A) and NGC~1316 was indicated
by earlier studies of the inner dust distribution and the outer system
of shells and loops (Schweizer 1980).  NGC~5128 is considered to be
the prototype remnant of a merger of two disks. NGC~1316 has also been
considered as a merger remnant (Bosma \etal\ 1985) from its high
luminosity and low central velocity dispersion compared to that
expected from the Faber-Jackson relation (D'Onofrio \etal\ 1995).
Interestingly, both galaxies' luminosity profiles follow an $r^{1/4}$
law, and show an outer elongated feature orthogonal to the inner dust
lane. This elongated feature is evident in D.~Malin's picture for
NGC~5128 (Malin 1981) and in Figure~8 of Schweizer (1980) for NGC 1316, 
reproduced here as Fig.~\ref{schweizerplate}.  These morphological
facts are consistent with incomplete violent relaxation in a
relatively recent merger.

The rapid rotation of the PNe system in the outer regions of
NGC~5128 is associated with the elongated light distribution (Hui
\etal\ 1995).  From numerical simulations of mergers between identical
disk--bulge--halo galaxies (Barnes 1992, Hernquist 1993) one expects
to see the angular momentum of the merger remnant to be
concentrated in its outer regions. Is this prediction borne out by the
kinematics of NGC 1316?  To test the similarity with NGC~5128 and in
particular the segregation of angular momenta in NGC~1316, we measured
radial velocities of 43 PNe which were originally identified by
McMillan \etal\ (1993).  The PNe spectra were obtained with the ESO
NTT, and the EMMI spectrograph.

We combine the information from this radial velocity sample at large
radii with the deep integrated-light absorption line spectra along the 
major and minor axis of NGC 1316, obtained at the Siding Spring Observatory 
2.3 meter telescope with the Double Beam Spectrograph (DBS). 
Our aim is to obtain as complete a description as possible of the kinematics 
of this galaxy. The analysis of the streaming velocity field is carried 
out by a non-parametric smoothing algorithm due to Wahba \& Wendelberger 
(1980), which makes no assumption of any prior functional form for the 
velocity field. In this regard it is similar to the work of Tremblay 
\etal\ (1995), who constructed a smoothed velocity field from a sample of 
68 PNe radial velocities in NGC~3384 and estimated the total mass of that
galaxy. Our analysis is different from theirs in the way in which the
degree of smoothing is determined (Tremblay \etal\ give no indications
as to the criteria adopted for this), and in the way symmetries are
used to determine the line of maximum gradient.  In view of the
still small number of PNe velocities we regard our study as a case
study, which will only find its full application when much larger samples
of PNe radial velocities become available from the 8 meter
telescopes.

We present the spectroscopic observations of the PN $\lambda 5007$
\oiii\ emission and the absorption line spectra in Section~2.  We then
address in detail the properties of the PNe velocity field in
Section~3.  In Section~4 we combine the PNe and integrated light data
to derive a global rotation field and velocity dispersion curve. In
Section 5 we consider the dynamical support of NGC 1316 
and estimate its mass from the Jeans equation. Conclusions and future 
developments are discussed in Section~6.

\section{Observations}

NGC 1316 (Fornax A) is a D galaxy in the Fornax cluster with a
disturbed outer morphology that is indicative of a merger $4 \times 10^8
- 2\times 10^9 \yr$ ago (Schweizer 1980). In this paper we take a distance 
of $16.9 \mpc$ (Mc Millan \etal\ 1993) so that $1''$ corresponds to $82\,\ppc$.

\subsection{Planetary Nebulae Spectra}

On December 22-24, 1995, we obtained the spectra for the \PN (PNe) in
the outer region of \n1316 ($\alpha (1950)\, =
03^{h}\,20^{m}\,47.0^{s} \,\, \delta (1950)\, = -37^\circ\, 23'\,
12''$) in the Fornax cluster.  We used the New Technology telescope
(NTT) EMMI spectrograph, in the red imaging and low dispersion mode,
with multi-object spectra (MOS) masks, the f/5.3 camera and the TEK
2048 CCD (24$\mu$m $= 0''.26$ pixel$^{-1}$), and the No. 5 grism: the
wavelength range is 4120$-$6330 \AA, with a dispersion of 1.3 \AA\
pixel$^{-1}$.  The only emission line visible in the spectra of these
faint PNe is \oiii\ $\lambda 5007$, so a filter with a central
wavelength $\lambda_c = 5050$ and FWHM $=500$ \AA\ was used in front
of the grism to reduce the wavelength range of each spectrum. This
allowed us to increase the number of slitlets in each MOS masks and
also to get enough emission lines from the calibration exposures for
an accurate wavelength calibration. The size of the slitlets punched
in the MOS masks is $1'' \times 7''$, which corresponds to $3.84
\times 27$ pixel at the CCD, and each mask covered a field of 
$5'\times8'$.

Astrometry and \oiii\ photometry for 105 PNe in \n1316\ came from
McMillan \etal\ 1993, and the PNe were identified from
on-band/off-band CCD images with the CTIO 4 meter telescope.  The
technique for producing MOS masks for targets which are not identified
directly from NTT EMMI images is discussed in detail by Arnaboldi
\etal\ 1994, 1996.  For the PNe in \n1316\ we used two masks: one was
offset $52''$ east from the center of \n1316, and the second was
offset $117''$ west. We produced a total of $70$  slitlets at the
positions of PNe in the two masks plus 3 slitlets on each mask at
positions of fiducial stars, to check the precise pointing of the
telescope. The total integration time was 3.8 hrs and 3.3 hrs for the east
and west mask respectively.

\begin{figure}\epsscale{.3}\plotone{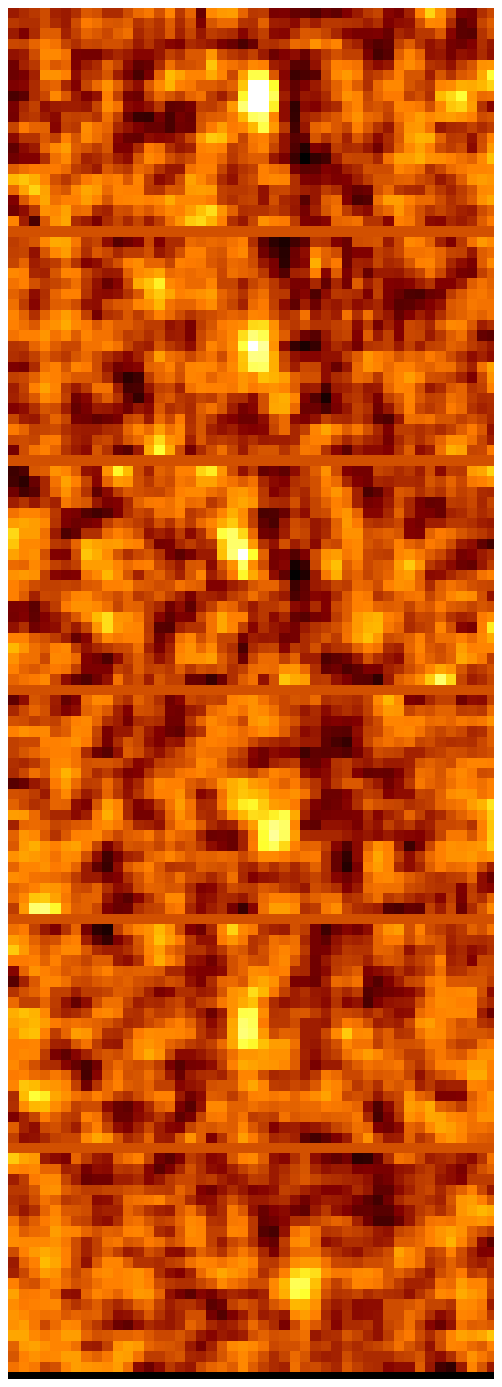}\caption{
Images of the 2-D spectra of some individual PNe in \n1316.
In each image the wavelength direction lies along the horizontal axis, and the 
spatial direction (i.e. the direction along the slitlet) lies along the 
vertical axis. Each spectrum shows about 61 \AA\ in wavelength. The spatial 
extent of the spectra varies but it can be estimated from the pixel scale of 
$0.26''$ per pixel.\label{fig1}
}
\end{figure}

\begin{figure}\epsscale{.7}\plotone{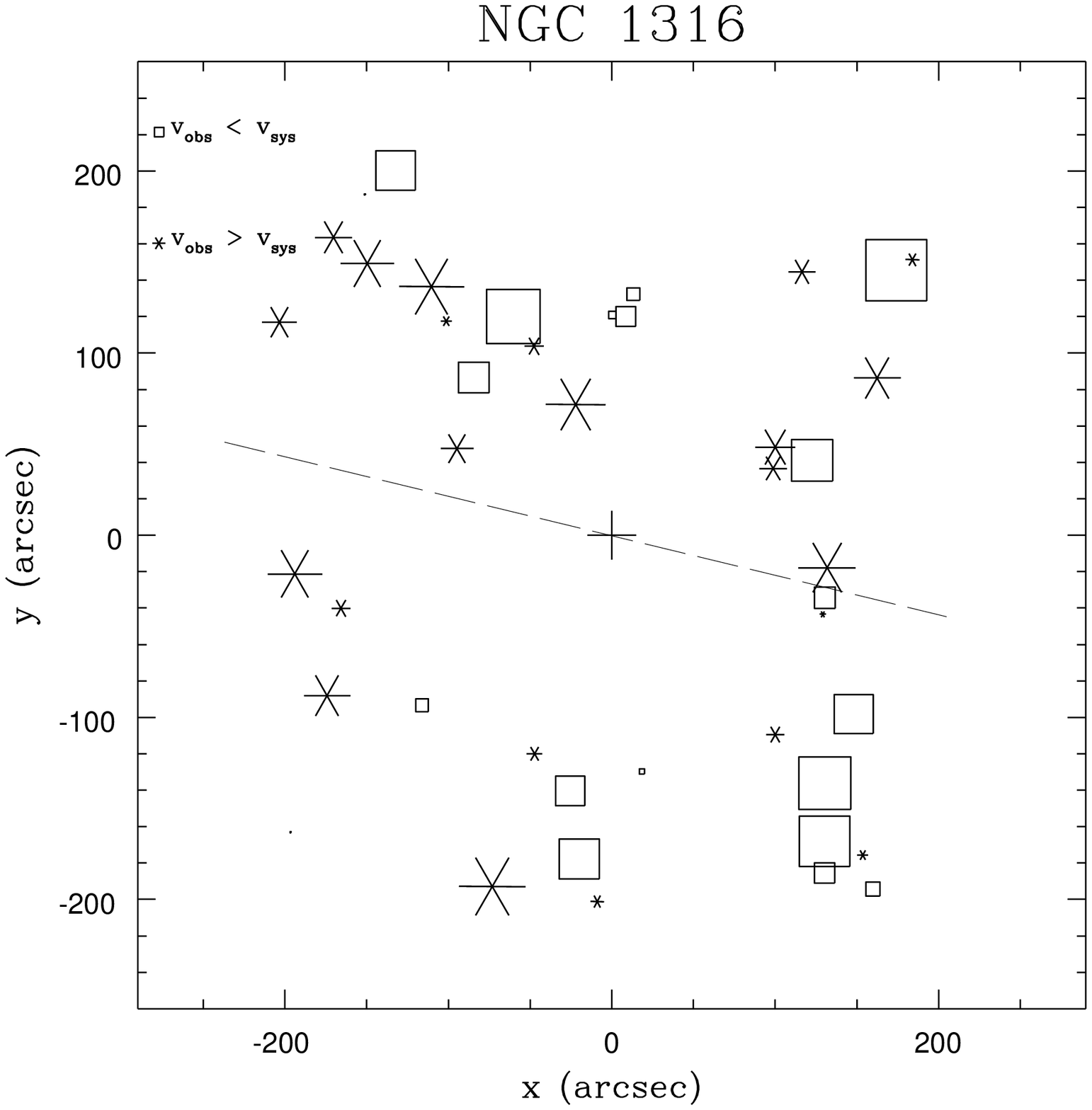}\caption{
The \PN spatial distribution. The symbol representing each nebula is
chosen so that open squares are for velocities approaching and crosses
for receding velocities; the symbol size is proportional to the
absolute value of the velocity with respect to the systemic
velocity. The thick dashed line shows the direction of maximum
gradient ($77^\circ \pm 28^\circ$) found from the linear least-square
fit. North is up, and East to the left.\label{f2pndistr}
}
\end{figure}

Flat-field exposures for the two MOS masks (east and west) were taken
with the internal lamp, the grism, and the interference filter. The
bias frames were flat and constant throughout the night.  The original
spectra were de-biased, flat-fielded, and registered for any small shifts
and rotation. Then the spectra for each mask were median-combined, and
the resulting frame appeared free of cosmic rays.  The 2-D spectrum
for each PN was extracted and wavelength calibrated using the
corresponding (He-Ar) arc spectrum. The PN spectra were then
individually sky-subtracted in two independent ways. In the first
method, the single PN spectrum was examined for a possible PN \oiii\
emission, and then an averaged sky was obtained from the part of the
spectrum free from the \oiii\ emission.  The resulting 1-D sky frame
was then smoothed, and the smoothed sky was subtracted from the 2-D PN
spectrum. In the second method, all the rows of the 2-D PN spectrum
were median-combined and the 1-D sky spectrum was fitted with a
low-order Legendre polynomial. This fitted sky continuum was then subtracted
from the 2-D PN spectrum. Images of 2-D spectra of some individual
PNe in \n1316\ are shown in Fig.~\ref{fig1}. 
The identification of the PN emission and the measurement of the central 
\oiii\ wavelength (via Gaussian fit) in the PNe spectra was done 
independently by R. Mendez and M. Arnaboldi. 
Agreement was found for 43 PN radial velocities in a
range of radii between 5~kpc and 22~kpc; the PNe velocity field
is shown in Fig~\ref{f2pndistr}. The velocity dispersion of
the PN radial velocity sample is $178 \pm 19 \kms$ and the average
velocity of the sample is $1783\,\kms$, in very good agreement with the
measured systemic velocity of \n1316 ($1793\, \kms$, from NED\footnote{
The NASA/IPAC Extragalactic Database (NED) is operated by the Jet propulsion 
Laboratory, Caltech, under contract with the National Aeronautic and Space 
Administration.}). The two MOS masks overlap and a number of PN appear on 
both masks: this gives a direct measurement of the error and any systematic 
shift in velocity between the east and west masks.  
13 PN slitlets were in common
between the East and West masks, and 7 PN spectra were actually
detected.  The mean of the distribution of the East-West velocity difference
$\Delta V $ for the detected \PN which are in common between the two masks is 
$0$, and the dispersion of $\Delta V$ is $60\, \kms$; the standard error on 
a single measure through one mask only is $42\, \kms$.

To check the reality of our detections, we looked for the weak \oiii\
$\lambda 4959$ emission from the PNe by averaging over all of the
candidate PN emission spectra, after shifting the spectra in
wavelength so that the \oiii\ emission lines were all at the same
apparent wavelength. In the averaged resulting frame for each mask
the \oiii\ $\lambda 4959$ line was well evident.

\newpage

\newpage

\subsection{Absorption line spectra}

Long-slit spectra were obtained for \n1316\ at the Siding Spring
Observatory 2.3 meter telescope with the DBS, simultaneously in the blue 
(4800-5800 \AA) and the red (8200-9200 \AA)
calcium triplet absorption line region. We used 1200 G/mm gratings on
both arms of the DBS.  The mean dispersion is 0.57 \AA\ pixel$^{-1}$
with a velocity resolution of $30\,\kms$. A Cu-He-Ar lamp was used to
wavelength-calibrate the spectra in the blue arm, while a Ne-Ar lamp
was used for the red arm of the DBS. The dimensions of the spectra are
$1850~(\lambda) \times 500$ pixels and the spatial pixel scale is $0.9''$
pixel$^{-1}$.

Long slit spectra were obtained along the galaxy major and minor
photometric axes, at ${\rm P.A.}= 50^\circ$ and $140^\circ$ respectively. 
Between each 2000s exposure on the galaxy, the telescope was moved to a 
nearby empty field where a corresponding sky spectrum was acquired, for 
an accurate subtraction of the sky lines. Calibration exposures were taken 
after each galaxy and sky frame, and internal quartz-lamp flats were obtained 
for both arms of the spectrograph.  
The total exposure time for spectra along ${\rm P.A.}= 50^\circ$ and 
$140^\circ$ was 1.6 hrs along each ${\rm P.A.}$. Spectra of standard stars, 
template stars and twilight sky were acquired at the beginning 
and at the end of the night. The mean seeing during the observations was
FWHM =$1.7''$.  The bias frames showed some transient structures,
which did not affect our measurements: bias subtraction was done using
the over-scan region at the edge of the CCD frames.

\begin{figure}\plotone{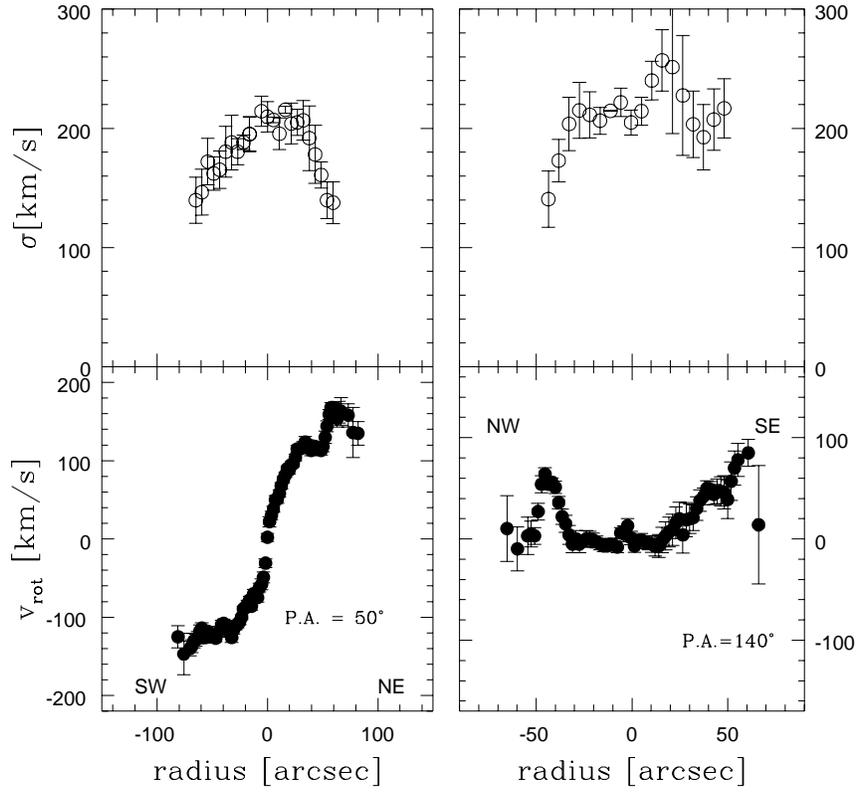}\caption{
Left: mean streaming velocity (lower panel) and velocity
dispersion measurements (upper panel) along the projected major (${\rm
P.A.} = 50^\circ$) axis. Right: same quantities for the minor
(${\rm P.A.} = 140^\circ$) axis.\label{f3absv}
}
\end{figure}

The long slit spectra were bias-subtracted, flat-fielded, and
wavelength-calibrated using the corresponding lamp exposures. The same
procedure was applied to the sky frames. For the sky subtraction, an
averaged sky frame was produced, and the cosmic rays were removed.
The sky continuum was fitted with a Legendre polynomial and subtracted
off.  Then the emission lines in the averaged-continuum subtracted sky
frame were normalized to the emission lines in each galaxy spectrum
and removed.  The resulting spectrum along each ${\rm P.A.}$ was obtained by a
median of all sky-subtracted spectra at each ${\rm P.A.}$.  The galaxy and
template spectra were normalized with respect to the stellar
continuum, and the radial velocity and velocity dispersion
measurements were obtained using the Fourier correlation quotient
method (Bender 1990). The rotation curve and the velocity dispersion
profiles are shown in Fig.~\ref{f3absv}.

\newpage

The spectra were filtered with an adaptive filter (Richter \etal\
1992), to reduce the noise in the outer regions, and velocities were
measured after binning two rows together, to match the seeing
conditions during observations. The Calcium triplet data at 8500 \AA\ agree 
well with the blue data in the inner regions, but do not reach out so far 
because of the bright sky lines.


\subsection{Summary of observational constraints}

\begin{figure}\plotone{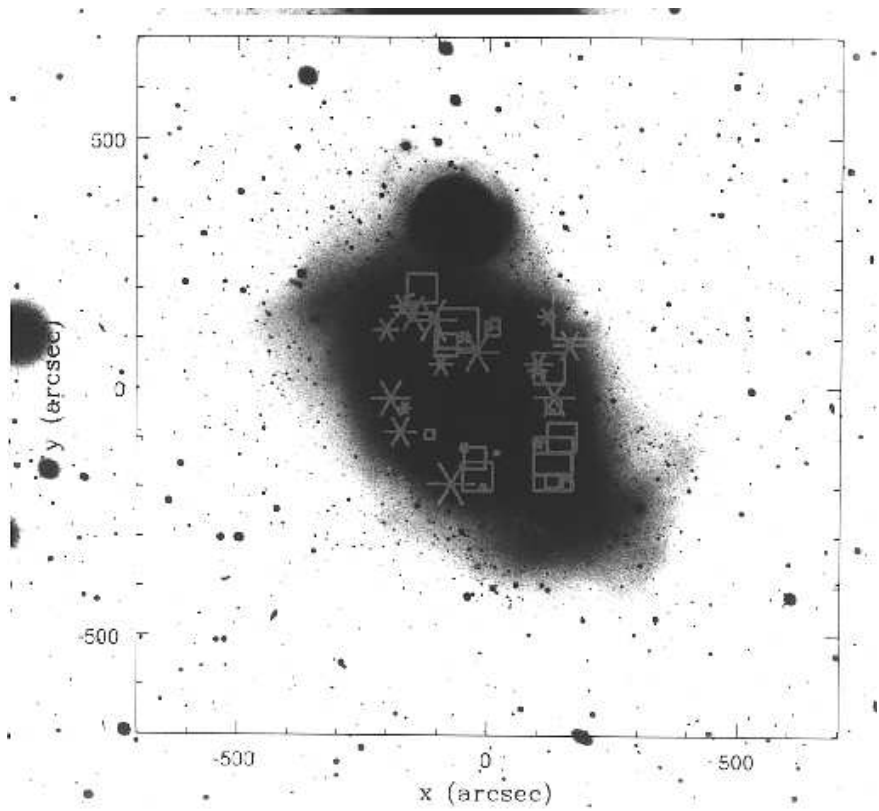}\caption{
Figure~8 from Schweizer's 1980 paper, with PNe radial velocity symbols.
\label{schweizerplate} 
}
\end{figure}

Images and photometry of \n1316\ were published by Schweizer (1980),
and Caon \etal\ (1994). The results of these papers can be summarized
as follows.  \n1316\ appears more disturbed than most elliptical
galaxies.  Prominent shells and plumes point to a recent merger
history. There are signs of a substantial dust component in the
central region of the galaxy, especially along the minor axis.  Inside
a major axis distance of $a=1'.5$, the image of \n1316\ is fairly
regular, like that of a normal elliptical galaxy. The position angle
of the major axis in this region is ${\rm P.A.} \simeq 58^\circ$, and
the ellipticity is about E3, but the isophote deviations from ellipses
are somewhat larger than for normal ellipticals.  The effective radii
along the major and minor axes are $a_e = 132''$ and $b_e = 90''$ respectively
(Caon \etal\ 1994), so $R_e = \sqrt{a_e b_e}= 109''$.  
Between $a=1'.5$ and $a=3'$, where the bulk of
our PN velocities are measured, the major axis position angle is
constant at ${\rm P.A.} \simeq 52^\circ$.  At $a\simeq 3'$ we identify
the first outer shells, and at that distance from the center a
prominent bar or thick disk starts to be visible, at
${\rm P.A.}=60^\circ$. At $a = 4'.5$ the position angle of the major axis is
at ${\rm P.A.} \simeq 54^\circ$. Fig.~\ref{schweizerplate} reproduces one of
Schweizer's (1980) deep images on which these structures can be seen.

\newpage

Despite the somewhat larger than normal isophote deviations inside
$3'$, no obvious asymmetry is seen in Schweizer's images in this inner
region. The photoelectrically measured brightness profile of \n1316\
approximately obeys the $r^{1/4}$ law from about $6''$ to at least
$6'$.  This and the approximate constancy of the major axis
position angle are consistent with the notion that the gravitational
potential of \n1316\ has largely settled down to a nearly triaxially
symmetric configuration, at least inside $3'$, but see below.

Is the PN sample representative of the underlying stellar population
in the outer region $(r> 90'')$ of \n1316?  The number of PN per unit
light in early-type galaxies depends on the integrated B--V of the
parent stellar population (Hui \etal\ 1993).  NGC 1316 does not seem
to have significant color gradient in its outer parts, where shells,
loops and tails become dominant. The photoelectric photometry by
Schweizer (1980) indicates a constant value in B--V$ \simeq 0.9$ from
$18''$ to $200''$. The aperture photometry from Lauberts \& Sadler (1984)
reports constant B--V $\simeq 0.9$ out to $300''$.  More recent results 
from Schr\"oder \& Visvanathan (1996) confirmed a constant color in B--V from 
$71''$ to $142''$. The existing multi-color photometry of NGC 1316 is
consistent with an homogeneous stellar population for the main
spheroid within $300''$. Hence we expect the true radial distribution
of PN in NGC 1316 to be proportional to the radial distribution of
integrated light.

However, the cumulative number density of the {\sl observed} PN sample
does not follow the underlying continuum light, except in the
outermost regions. We compared the cumulative number densities of the
PNe as well as of the continuum light as a function of surface
brightness. They are not linearly related inside $100''$, probably due
to selection difficulties of faint PNe against the galaxy continuum in
the central parts. In the following the PNe must therefore be regarded
as a test particle distribution, whose velocity distribution can be
regarded as locally representative of the true velocity distribution,
but whose density and phase-space density are effectively independent
of the true underlying stellar distributions.

The absorption line measurements described in Section 2.2 show that
the kinematic major axis is close to the photometric major axis: along
${\rm P.A.} = 50^\circ$, the radial velocities increase rapidly near the
center and then we find a streaming velocity plateau at $a \sim 100''$
of $v \simeq 145\,\kms$, similar to the value found by D'Onofrio \etal\
(1995) at $a=50''$ for the same ${\rm P.A.}$. Bosma \etal\ (1985) find a
plateau value of $v\simeq 140-150\,\kms$ at $80-100''$ along ${\rm P.A.}
= 60^\circ$, so the gradients away from ${\rm P.A.} = 50^\circ-60^\circ$ are
small.  The large rotation along the photometric major axis does not
seem to be related to a hidden disk component. Schweizer (1980) ruled
out substantial evidence for a disk component brighter than $B \sim
26$, and Caon \etal\ (1994) determined the $a_4$ parameter of the
isophotes inside $2'$ to be mostly negative, which is indicative
of boxiness.

The absorption line measurements plotted in Figure~\ref{f3absv} show that 
the kinematic minor axis of the stars within $70''$ must be near 
${\rm P.A.}= 140^\circ$, close to that of the photometric minor axis: 
there is no velocity gradient within $30'' $ of the galaxy nucleus. 
In this region there is no
evidence for a misalignment between photometric and kinematic
principal axes in the available absorption line data, and the stellar
velocity field is consistent with being axially symmetric.  At
larger distances along the galaxy minor axis, the radial velocities
increase on {\it both} sides, producing a {\bf U}-shape morphology in
the radial velocity curve, and the velocity dispersion profile is
irregular.  This kind of U-shaped velocity morphology is common among
interacting ellipticals (Borne \etal\ 1994), but in these
circumstances it lasts for only about one dynamical time. In NGC 1316 the
local dynamical time at $40''$ is $\sim 3\times 10^7 \yr$, while from
the relaxed morphological appearance inside $3'$, the merger cannot have
occurred later than $\sim 2\times 10^8 \yr$ ago. This, and the shape
of the NW part of the minor axis velocity curve, suggest that the
U-shaped velocity morphology is due to some late--infalling material.
If so, the U--shaped velocity curve is associated with a minor
perturbation which does not affect the global dynamics. This we will
assume in what follows.

\section{Analysis of the PN velocity field}

In this section we determine the properties of the PN velocity field
without reference to the absorption line kinematics. We will first, in
Section 3.1, fit a linear velocity field (equivalent to a solid body
rotation) and a flat rotation curve to the measured radial
velocities. We then investigate non-parametrically whether the
direction of maximum rotational gradient obtained from these fits is
sensitive to the assumed functional form of the velocity field
(Section~3.2).

\subsection{Parametric velocity fields}\label{section3}

We first make a linear fit to the original sample of 43
radial velocities $v_{rad} (x, y)$ of the form 
$$ v_{rad} = a_0 x + a_1 y +c. $$
 Here $x$ and $y$ are pixel coordinates on the CCD frame. $a_0, a_1,
c$ are determined by a least square fit to the observed radial
velocities $v_{rad}$. This linear velocity field is equivalent to
solid body rotation of the form
$$
 v_{rad} = v_{sys} + \omega r \cos(\varphi-\varphi_0). 
$$ 
where $\varphi$ is the position angle of the PN, $\varphi_0$ is the position
angle of the kinematic major axis, $r$ is the radius in arcsec, 
and $v_{sys}$ is the systemic velocity. The origin is the
mean $(\bar{x}_{PN}, \bar{y}_{PN})$ for the sample.
This will give a first indication of whether mean rotation is present
in the outer parts of \n1316.  This linear fit gives the following
parameters: $\bar{x}_{PN} = 1212,\, \bar{y}_{PN} = 1016, \,
v_{sys} = 1783\pm 26\, \kms$ with a maximum gradient of $0.11 \pm
0.05\, \kms$ pixel$^{-1} = 0.42 \pm 0.19\, \kms$ arcsec$^{-1}$ along
$\varphi_0 = {\rm P.A.} = 77^\circ \pm 28^\circ$ (measured from North
towards East).  This direction is shown in Fig.~\ref{f2pndistr}; along
this ${\rm P.A.}$ the mean rotation is $100 \pm 45\, \kms$ at radius of 
$4' = 19.7$ kpc. Once this linear component is subtracted from the \PN radial
velocities, the dispersion of the velocity residuals is $170 \pm 18\,\kms$. 
We note that this value is very similar to the absorption
line velocity dispersion of $\sigma\approx 160\, \kms$ at $60''$.
The centroid of the PNe resulting from the linear fit is within
$7''$ of the photometric center of NGC 1316.

We also fitted a flat rotation curve directly to the 43 PN radial
velocities, using the following function:
$$
 v_{rad} = v_{sys} + v_o\cos(\varphi-\varphi_0).
$$
The equations that give the three parameters were solved numerically.  
The errors on the parameters were computed from the partial variation of 
$\chi^2$ with each parameter.  For the flat rotation curve fit, we get the 
following values for the parameters: $v_{sys} = 1783\pm 27\, \kms$, 
$\varphi_0 = 70^\circ \pm 35^\circ$, $v_0 = 67\pm 37\, \kms$, and the velocity
dispersion is $\sigma = 175\, \kms$.  The parameters $\varphi_0,\, v_0$
for the flat rotation curve fit are marginally worse determined than
$\varphi_0,\, \omega$ for the solid body rotation fit. The residuals
showed no radial trends in either case.

\subsection{Non-parametric fitting of the velocity field}

In this section we investigate the properties of the 2-D PN velocity
field without assuming an ``a priori'' form for the rotation curve or
velocity field. The sample of 43 radial velocities is rather small for
fitting a non-parametric velocity field. However, based on the
discussion in Section 2.3 we assume that the velocity field is
point-symmetric with respect to the center of the galaxy, i.e., that
to every velocity $v$ at position $(x,y)$ on the sky there corresponds
a velocity $-v$ at position $(-x,-y)$. Such an assumption would be
justified both in the case of a general velocity field in a triaxially
symmetric potential (resulting in a projected velocity field with
minor axis tilt) and for an axisymmetric velocity field (resulting in a
bisymmetric projected velocity field without tilt). Assuming point
symmetry we construct a sample of 86 data points by
reflecting all observed data points around the galaxy center.  By
fitting a two-dimensional surface to these 86 velocities we derive a
smoothed velocity field $\bar{v} (x,y)$.

\subsubsection{The smoothing algorithm}

We have used the non-parametric smoothing algorithm of Wahba \&
Wendelberger (1980), originally developed for analyzing meteorological
data, to determine the properties of the velocity field
$\bar{v}(x,y)$.
This algorithm uses thin plate splines to represent a smooth
solution $\Phi(\vec x)$ which minimizes the quantity:
$$
 \frac 1N \sum_{i=1}^N \left [ \Phi(\vec x_i) - \Phi_i \right ]^2
         + \lambda \int_{-\infty}^\infty
   \left (
       \left [ \frac{\partial^2 \Phi}{\partial x^2} \right]^2
     + 2 \left [ \frac{\partial^2 \Phi}{\partial x \partial y} \right]^2
     + \left [ \frac{\partial^2 \Phi}{\partial y^2} \right]^2
   \right )
dxdy.
$$

Here $N$ is the number of observations $\Phi_i$ at positions $\vec x_i$.
The degree of smoothness is controlled by the
regularization parameter $\lambda$; in the following, we use
$\Lambda=\log (N\lambda)$.

For large values of $\Lambda$, the algorithm in this form will fit a
plane through the measured points. For small $\Lambda$ it will
minimise deviations between model and data, even at the expense of a
noisy model.  There will then be a range of $\Lambda$ for which an
adequately smooth, non--parametric solution is found which at the same
time is a fair representation of the data. For sufficiently good data,
the surface from which the data are drawn will be recovered.

The present application is somewhat different from the
standard case in that, even for zero observational errors, the
measured PNe radial velocities would locally not agree with the fitted
velocity field at their positions, due to the galaxy--intrinsic
dispersion $\sigma$ in the PNe velocities.  In fact, the velocity
dispersion dominates over the observational errors. Thus in our case
the algorithm will find the best-fitting smooth streaming velocity
field by minimising $\sigma$ as much as possible within the imposed
smoothness constraints. We therefore need to discuss carefully
how the imposed degree of smoothing is chosen.

\subsubsection{Determining $\Lambda$.}

In the standard case of fitting a function to a sample of measurements
drawn from an underlying smooth surface, we can distinguish the
following two extreme regimes: (i) For large $\Lambda$ the algorithm
would fit a plane through the measured points. If the underlying
surface from which the data points are drawn is poorly described by a
plane, and the measurement errors are sufficiently small, this would
be recognisable in that the mean $\chi^2$ per data point with respect
to the fitted surface, $<\chi^2>$, would be large.  (ii) On the other
hand, for small values of $\Lambda$ the algorithm would try to fit a
very un-smooth velocity field going through all points exactly, with
the result that $<\chi^2>$ would be much less than one.

A simple way of determining an appropriate value for $\Lambda$ would
then be to require that the $<\chi^2>$ with respect to the fitted
surface should be about unity.  For such a value of $\Lambda$ the
fitted surface would be consistent with the measurements, within their
errors, but it would not be unduly noisy, because the random
fluctuations in the measurements around the underlying true values are
not followed in detail.  In our case, for small $\Lambda$, the
algorithm would again try to fit all measured points exactly, except
that this is harder because of the large dispersion between
neighbouring points. For large $\Lambda$, however, the situation here
is different. Since the mean streaming velocity is nowhere larger than
the velocity dispersion $\sigma$, even for a velocity field rising
non-linearly with $\vec{x}$, $<\chi^2>$ could never become much larger
than unity. Thus determining $\Lambda$ from the condition that
$<\chi^2>\simeq 1$, while still possible, would not lead to a very
well-determined value of $\Lambda$.  Moreover, we have seen above that
the 43 measured PN velocities in \n1316\ are consistent with a nearly
linear velocity field.  Therefore, since we will want to determine the
velocity dispersion $\sigma$ at the same time as the best value for
$\Lambda$, we can not use it to fix $\Lambda$.

An often--used procedure to determine the smoothing parameter is the
so--called generalised cross validation method (GCV; Wahba \&
Wendelberger 1980).  The idea of cross validation (CV) is to use only
$N-1$ rather than $N$ data points to determine the function, ask how
well this then predicts the $N^{th}$ measurement, repeat this
procedure $N$ times cutting out each of the data points once, evaluate
a $<\chi^2>$, and choose $\Lambda$ such that this $<\chi^2>$ is
minimised. GCV is a version of CV that uses a fast approximation for
the cross validation function, making the method much easier to
apply. In addition, assuming uncorrelated residuals, CV schemes can be
used to estimate the mean variance.

Unfortunately, however, GCV gives a robust estimate for $\Lambda$ only
when one has at least $\sim 15$ data points with reasonable S/N (in
our case, in the dispersion; cf.\ Wahba \& Wendelberger 1980). In
fact, we have tried GCV on our small sample, which corresponds to
about three such data points, and have found that the method prefers
infinite $\Lambda$. This means that GCV cannot distinguish the data
from a plane.

Instead we use a Monte--Carlo method to determine $\Lambda$. The
idea is to draw artifical data from a known velocity field
and see how well one can recover this velocity field from the
data in a statistical sense, as a function of $\Lambda$. From the
discussion above, one expects the recovered velocity field to
deviate from the input field for both small and large $\Lambda$.
Both the characteristics of the data and the input velocity field
should be as close to the real case as possible.

Thus we construct a characteristic bisymmetric velocity field $v_{\rm
in}$, fitting to the observed PNe velocities with $\Lambda=2.2$ (a
plausible value).  We then draw samples of artifical data from $v_{\rm
in}$, and analyse them exactly as we will analyse the
observations. Each sample consists of 43 randomly drawn radial
velocities at the positions of the observed PNe. The velocity
distribution at each position is a Gaussian with dispersion
$\sigma=165\kms$ centered on $v_{\rm in}$.  Then, for a range of
values of $\Lambda$, we determine the line of maximum gradient as
described in section~\ref{lomg}, and fit a bisymmetric velocity field
$v_{\rm MC,\Lambda}$ to the 172 bi--symmetrized artificial data
points, using the particular $\Lambda$ and the line of maximum
gradient obtained with this $\Lambda$.  On a grid in ($x,y$) we
finally determine the quantity $\chi_{\rm MC,\Lambda}^2 = [ v_{\rm
MC,\Lambda} - v_{\rm in} ]^2 $, which gives the average quadratic
deviation of the velocity field determined by the data, from the
velocity field from which the data were drawn.  Repeating this for a
large number of Monte Carlo samples results in a smooth average
$\chi_{\rm MC}^2(\Lambda)$, which should have a minimum at a certain
value of $\Lambda$. In a statistical sense, a sample of 43 PNe is best
analyzed with this optimal $\Lambda$, and therefore this value should
be used in the fitting algorithm.

\begin{figure}\plotone{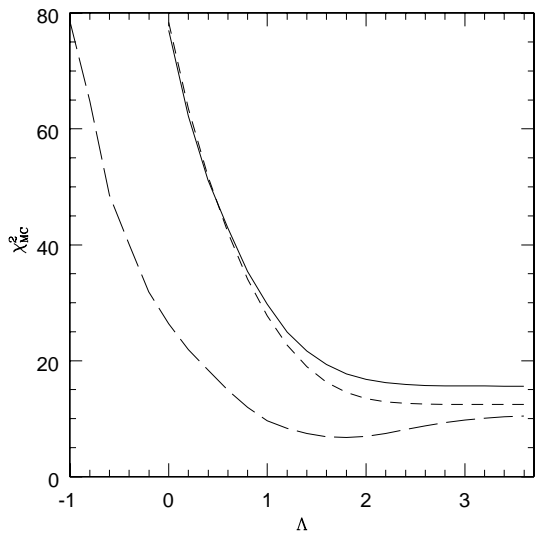}\caption{\label{figMC}
Average deviation $\chi_{\rm MC}^2(\Lambda)$ of an inferred
bisymmetric velocity field from the input velocity field $v_{\rm in}$,
as a function of $\Lambda$.  The solid curve shows the results for a
field $v_{\rm in}$ determined from the 43 observed PNe velocities
using $\Lambda=2.2$ and a velocity dispersion of $\sigma=165\kms$. The
short--dashed line is for 43 PNe drawn at random positions from an
$\arctan$ velocity field with the same dispersion.  The long--dashed
line is for the same $v_{\rm in}$, but with a sample of 100 PNe. A
clear minimum is visible only for 100 data points.}
\end{figure}

Fig.~\ref{figMC} shows $\chi_{\rm MC}^2(\Lambda)$ (solid line) for 450
samples of 43 velocities each, drawn from $v_{\rm in}$. As expected,
small values of $\Lambda$ lead to large errors in the reconstruction
of $v_{\rm in}$, due to amplified noise. However, because of the
near--linearity of $v_{\rm in}$, the small number of observed radial
velocities, and the large ratio of dispersion to rotation velocity, no
increase towards large $\Lambda$ is visible.  For $\Lambda \gta 2$,
$\chi_{\rm MC}^2(\Lambda)$ is nearly constant which means that with 43
velocities and $\sigma=165\kms$ it is not possible to distiguish
$v_{\rm in}$ from a planar velocity field -- all velocity fields
obtained for $\Lambda \gta 2$ are equally plausible.  The same is
true for a cylindrical velocity field with more spatial structure,
drawn from an $\mbox{arctan}$-function: then also no minimum is found
(short dashed line).  However, increasing the number of points to 100
in this case results in a clear minimum (long dashed line). In the
following, we use as a plausible value for the smoothing parameter the
smallest $\Lambda$ for which $\chi_{\rm MC}^2(\Lambda)$ has not
increased significantly, i.e., $\Lambda=2.2$. This ensures that the
regularisation suppresses as little structure in the velocity field as
possible, while at the same time the small number of data points is
not overinterpreted.

The smoothed velocity field $\bar{v}(x,y)$ obtained by fitting a 
two-dimensional surface to the velocities of the point-symmetric PN
radial velocity sample is shown in Fig.~\ref{f6vfield}.
Fig.~\ref{f5hisgauss} shows the residual velocity distribution for 
this case. The width $(\Delta v)_{\rms}=165\, \kms$ of this distribution is 
our best estimate for the PN velocity dispersion $\sigma$.


\subsubsection{Determining the direction of maximum rotational gradient}
\label{lomg}

Now we derive the properties of the velocity field $\bar{v} (x,y)$. We
first want to determine the angle $\varphi$ for the direction of
maximum gradient, independent of the functional form adopted for the
velocity field. To this end, for any value of
$\varphi$ in the range $[0,180]$ degrees,

(1) we symmetrize the 86 velocity data points around an axis which is
rotated by $\varphi$ with respect to the $x$-axis of the CCD frame
(centered on \n1316). From the set of 86 data points we thus
obtain 172 data points.  If $x_{PN}, y_{PN}$ are the PN coordinates in
the CCD frame, the symmetric data points are derived from the measured
PN coordinates as follows:
\begin{eqnarray*}
x'_{PN} = x_{PN} \cos\varphi  + y_{PN} \sin\varphi   &\rightarrow&
	x'_{\rm sym} =  x'_{PN} \\
y'_{PN} = -x_{PN} \sin\varphi  + y_{PN} \cos\varphi  &\rightarrow&
	y'_{\rm sym} = -y'_{PN} \\
v'_{\rm rad}= v _{\rm rad} &\rightarrow& v'_{\rm sym} = -v'_{\rm rad}
\end{eqnarray*}

(2) we fit a bisymmetric velocity field $\bar {v}_2 (x',y'|\varphi)$
to these 172 symmetrized data points, using the Wahba \&
Wendelberger algorithm and a smoothing parameter $\Lambda = 2.2$.

Then we determine $\varphi$ by minimising $\chi^2 \equiv [\bar{v}_2
(x',y'|\varphi) - \bar{v}(x,y)]^2$ on a grid in the
$(x',y')$--plane. This results in the best bisymmetric approximation
to the original velocity field $\bar{v}(x,y)$. Fig.~\ref{f7detphi}
shows $\chi^2 = [\bar{v}_2(x',y'|\varphi) - \bar{v}(x,y)]^2$ as a
function of $\varphi$.  Note that there is a distinct minimum at
$\varphi_{\rm min}=80^\circ$, at which the rms deviation between the
two smoothed velocity fields $\bar{v}(x,y)$ and
$\bar{v}_2(x',y'|\varphi)$ is $6 \,\kms$.  Thus at the high level of
smoothing required by the small number of PN velocities, the smoothed
velocity field $\bar{v}$ is consistent with four--fold symmetry.
There is no significant dependence of $\varphi_{\rm min}$ on
$\Lambda$; even in the case of the noisy velocity fields obtained for
$\Lambda \simeq 1.5$ the best $\varphi$ differs from $\varphi_{\rm
min}$ only by $\sim 10^\circ$.  The fact that the minimum is well
determined implies that there is a significant direction of maximum
gradient.

We determine an error on $\varphi_{\rm min}$ in two ways. First, we
compute the likelihood of the 43 measured PN velocities, given a
velocity field $\bar{v}_2(x,y|\varphi)$:
$$
\log P_{\rm PN} (\varphi) = - \sum_i (\bar{v}_2(x,y|\varphi)-v_i)^2/2\sigma^2
			+ {\rm const.,}
$$
where the position and radial velocity of planetary nebula $i$ are
denoted by $x_i$, $y_i$, $v_i$, and the velocity dispersion is
$\sigma=165 \,\kms$, as determined for $\Lambda=2.2$ and consistent with
the absorption line velocities in Fig.~\ref{f3absv}. The relative
likelihood $P_{\rm PN}(\varphi) / \mbox{max} P_{\rm PN}(\varphi)$ is
plotted in Fig.~\ref{f7detphi}.  Relative to the maximum likelihood
value of $\varphi=76^\circ$, the function has decreased to
$\exp(-1/2)$ at $\varphi=(109^\circ,48^\circ)$.  Second, we have
determined the distribution of $\varphi_{\rm min}$ for a set of Monte
Carlo generated data sets of 43 data points each, drawn from the
point--symmetric velocity field $\bar v(x,y)$. This has a maximum at
$\varphi_{\rm min}=81^\circ$ with a dispersion of $30^\circ$. Both
estimates of the error are consistent.  From the non-parametric fit of
the 43 PNe radial velocity sample, we conclude that the line of
maximum gradient for the PN data (hereafter LOMG) is $\varphi=80^\circ
\pm 30^\circ$.  This inferred direction is in good agreement with the
${\rm P.A.}$ determined via the parametric function fitting.


\subsubsection{Properties of the velocity field}

\begin{figure}\plotone{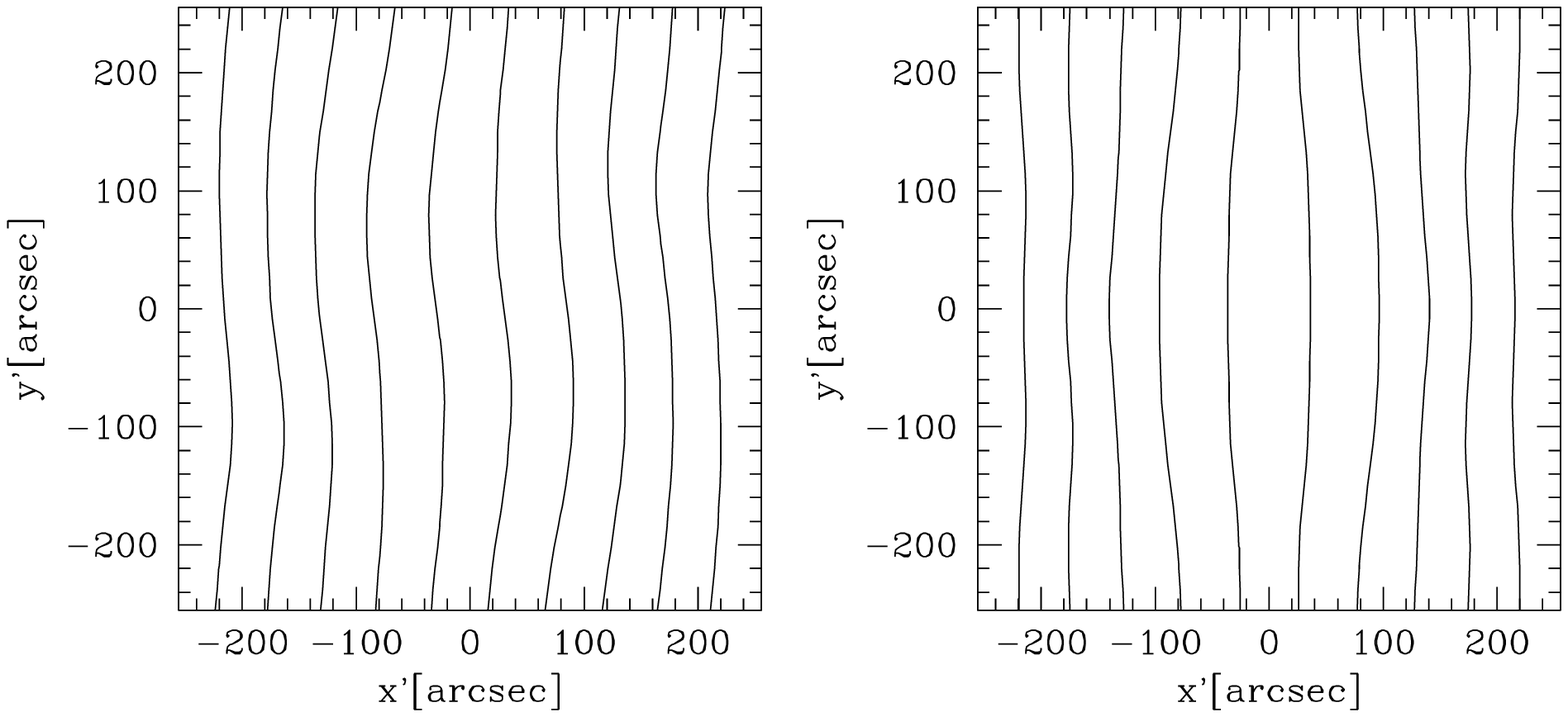}\caption{
Left panel: smoothed point-symmetric velocity field $\bar{v}(x',y')$ for
$\Lambda=2.2$, rotated to coordinates where the $x'$--axis is at
position angle $\varphi = 80^\circ$. Right panel: four--fold symmetric 
velocity field $\bar{v}_2(x',y'|\varphi)$ for the optimal major--axis direction
$\varphi_{\rm min} = 80^\circ$ and $\Lambda=2.2$. The contours are
spaced by $\Delta v= 22.7~\kms$, decreasing from left to right, and the
central contour has $v=9~\kms$.\label{f6vfield} 
}
\end{figure}

Fig.~\ref{f6vfield} shows the derived PN velocity field rotated such that the
$x$-axis lies along the LOMG at $\varphi= 80^\circ$. For comparison, we
also show the bisymmetric velocity field $\bar{v}_2(x',y'|\varphi)$;
the difference between these is not significant (\rms\ $6\,\kms$).  The
rotation of about $140\, \kms$ at $4'$ is clearly apparent.

\begin{figure}\plotone{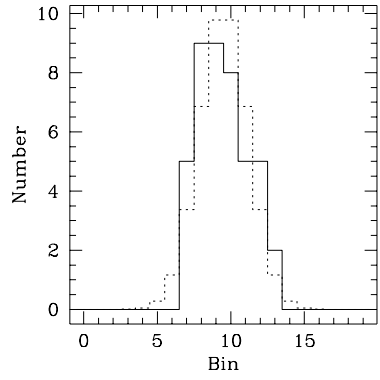}\caption{
The distribution of residual velocities from the fitted
velocity field $\bar{v}$ for $\Lambda=2.2$ (solid line) and a Gaussian
velocity distribution with the same dispersion $\sigma=165\, \kms$
(dotted line). The bin size is $100\, \kms$.\label{f5hisgauss}
}
\end{figure}

\begin{figure}\plotone{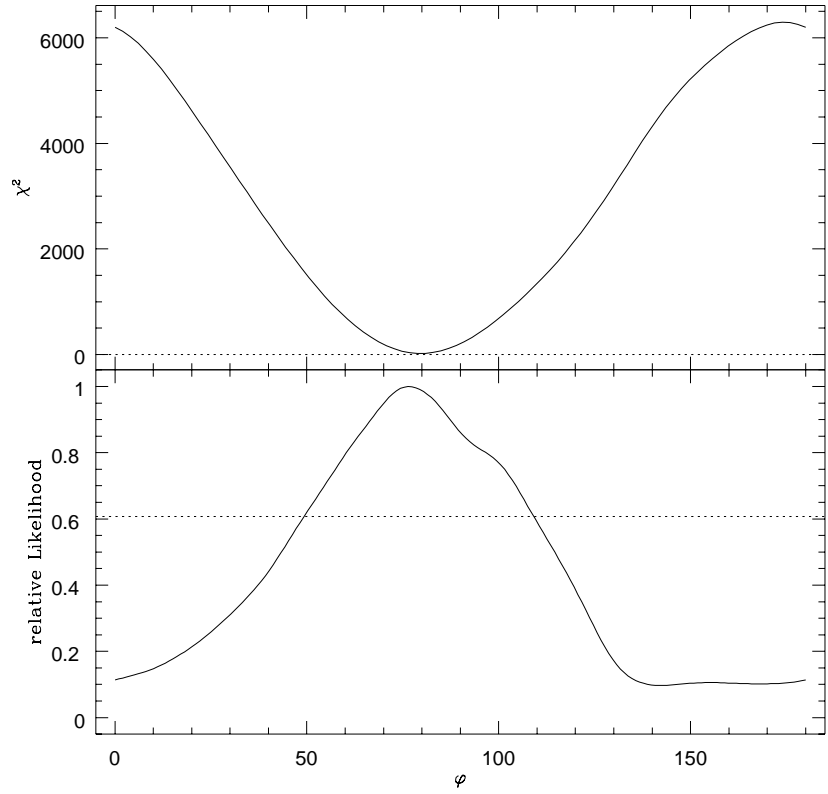}\caption{
Top: $\chi^2 = (\bar{v}_2(x',y'|\varphi) - \bar{v}(x,y))^2$ as
a function of $\varphi$. Bottom: relative likelihood of the
measured PN velocities for $\bar{v}_2(x',y'|\varphi)$ as
a function of $\varphi$. \label{f7detphi}
} 
\end{figure}

The residual velocity distribution about this velocity field is nearly
Gaussian, as already shown in Fig.~\ref{f5hisgauss}. The value of
$\chi^2 = [\bar{v}_2(x',y'|80^{\circ}) - v_{\rm rad}]^2 = 2.73 \times
10^4 (\kms)^2$ which gives a $\sigma _{PN} = 165\,\kms$ after the mean
streaming velocity field is accounted for. This is similar to the
values derived from the parametric fits.

For our best value of $\varphi_{\rm min}=80^\circ$, we project the
point symmetric data set of 86 points onto the LOMG, and then compute
a smoothed rotation curve via Wahba \& Wendelberger's smoothing algorithm.  
For small $\Lambda$, the resulting rotation curve shows substantial
counter-rotation in the central parts, which is excluded by the
absorption line data and must therefore be an artifact of the small
number of PN velocities in the sample. Thus we take for $\Lambda$ the
minimum value for which we do not have counter rotation in the central
region ($\Lambda =4$).  By choosing $\Lambda$ in this way, we also
smooth away some real velocity gradient in the outer parts.

\begin{figure}\plotone{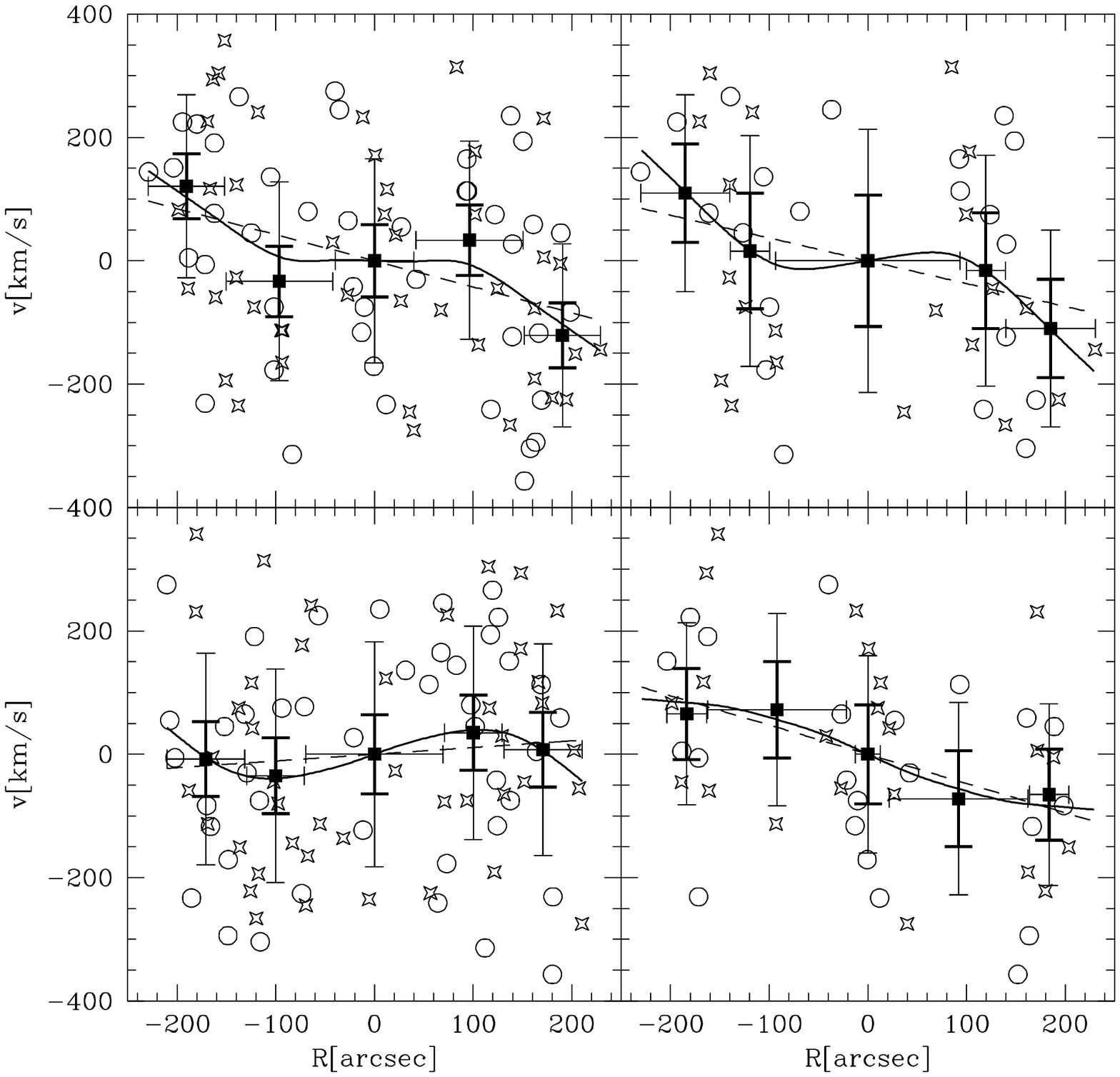}\caption{
Smoothed rotation curves from PN velocities.  Top left: along the
LOMG, for the point symmetric data set 86 velocities (with $\Lambda = 4$). 
Different symbols are used to identify the actually observed 43
points (open circles), and their symmetric counterpart (stars).
Bottom left: same, along the axis orthogonal to the LOMG.  Top right:
along the LOMG, but only for data points at heights less than
$121.5''$ above the LOMG.  Bottom right: same, but only for data
points at heights larger than $121.5''$ above the LOMG.  In each panel
the full line is obtained with $\Lambda = 4$, the dashed line shows
the linear fit from Section~3.1.  The squares and thick error bars
indicate the mean velocity and its error, the thin ``error bars'' the
calculated velocity dispersion $\sigma$, all computed in a small
number of bins each containing about one fifth of the PNe.  The
$1\sigma $ error on
the mean velocity is given by the velocity dispersion divided by
$\sqrt{N_{\rm bin}}$, i.e., about $60 \kms$, making the outermost
rotation a $2.3\sigma$ result.
\label{f8rotcurve}
}
\end{figure}

Fig.~\ref{f8rotcurve} shows the smoothed rotation curves so derived
along the LOMG ($\varphi=80^\circ$) and the perpendicular direction
($\varphi=170^\circ$). No significant rotation is seen along the
perpendicular axis.  Along the LOMG there is clear rotation in the
outer parts and we obtain a rotation of $140 \pm 60\, \kms (1\sigma)$ at
$4' = 19.7$ kpc. It is not clear whether we have rotation in the inner
parts.  This is because the PNe sample lacks objects at small
distances from the galaxy center, and the radial velocities of PNe
which appear at small projected distances from the center in
Fig.~\ref{f8rotcurve} actually sample the gravitational potential high
above the major axis. Therefore the mean velocity curve shown in the
top left panel of Fig.~\ref{f8rotcurve} really provides no information
about rotation at radii less than $80''$ from the center. In the next
section, we will use the absorption line data to fill in this gap.

We next divide the PN velocity points into two about equal subsamples:
those points whose $y' < 121.5'' = 9.8 $ kpc from the LOMG, and those
for which $y' > 121.5''$. We project both sets of velocities
independently onto the LOMG and fit a rotation curve to each set
separately with the same $\Lambda =4$. Fig.~\ref{f8rotcurve} shows at
best a hint of the PNe with $y' < 121.5''$ rotating faster at the larger
radii ( $179\, \kms$ at 18.5 kpc ) than those with $y'> 121.5''$. The
fast rotation signal at large major axis radius near the plane comes
from PNe on both sides of the center.  The difference  between both subsamples
is only $\simeq 0.5 \sigma$.  Also, if we fit a linear rotation curve to both
subsets of the data, the derived slopes are nearly identical. Thus any
non-cylindrical rotation needs to be confirmed with a larger sample of
PN velocities.

Inspection of Fig.~\ref{schweizerplate} reproduced from Schweizer (1980)
shows an elongated bar-like or thick disk structure at a distance of
$\sim 3'$ from the center of \n1316\ along a ${\rm P.A.} = 60^\circ$. The
large rotation velocity in the outer parts inferred from the PN
velocities might reflect rotation connected with this structure, which
is within $1\, \sigma$ of the ${\rm P.A.}$ of the LOMG. This is a similar
situation to the outer region of Cen~A, NGC~5128, where the outer PNe
with high rotation and small $\sigma$ are associated with an elongated 
light distribution.

\newpage

\newpage

\newpage

\newpage

\section{Combined velocity field from PNe and absorption line data}

\begin{figure}\plotone{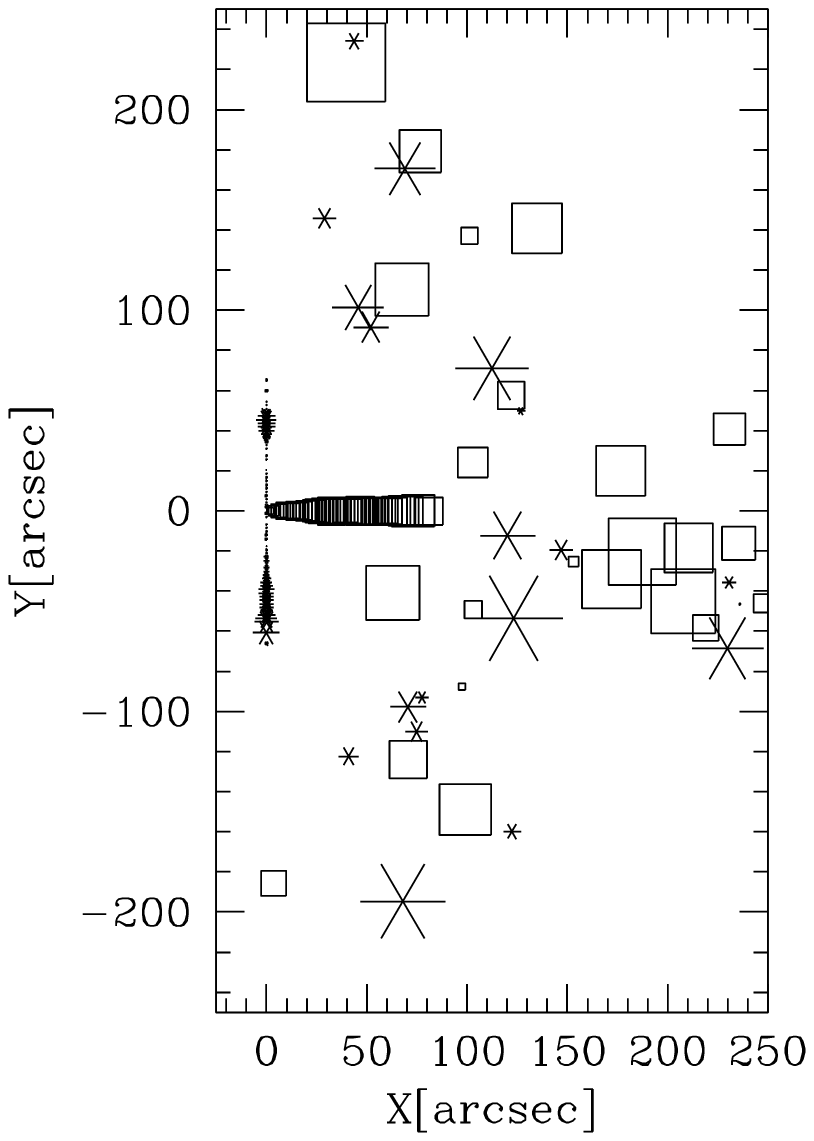}\caption{
Rotation velocities in the combined data. The densely packed points
along $y=0$ represent the major axis rotation velocities from the
integrated light data along ${\rm P.A.}=50^\circ$. Those along $x=0$ represent
the minor axis velocities along ${\rm P.A.}=140^\circ$. The isolated symbols
show the velocities for the individual PNe. Those at negative major
axis coordinate $x$ have been reflected about ${\rm P.A.}=140^\circ$ and their
velocities reversed. The coding is the same as in Fig.~\ref{f2pndistr}
(open squares are for velocities approaching, size of the symbol
proportional to absolute velocity). \label{verteilung}
}
\end{figure}

The PNe provide us with little information about the rotation
near the center of the system. We therefore now construct a
velocity field using the integrated light spectroscopy in the inner
regions and the PNe velocities in the outer regions. Fig.~\ref{verteilung}
shows the combined data with the PNe reflected about the adopted minor axis
of the absorption line data, reversing their sign in velocity when
reflected.

To the combined data, we fit a velocity field using the smoothing
algorithm, with $\Lambda=1.2$ to resolve the stronger curvature in the
absorption line data.  Here the relative weights of the PN absorption
data and the PNe were chosen according to their measurement
errors. The error for the PNe velocities is $60 \kms$; the errors for
the absorption line data are shown in Fig.~\ref{f3absv}. Where there
are absorption line velocities, they essentially fix the fitted curve;
where there are none they only enter through the regularisation. The
PNe fix the curve far from where absorption line data exist, but have
almost no effect in regions where the absorption line data
dominate. Giving the PNe data additionally different weights results
in a very similar curve when $\Lambda$ is changed according to the new
weights.

\begin{figure}\plotone{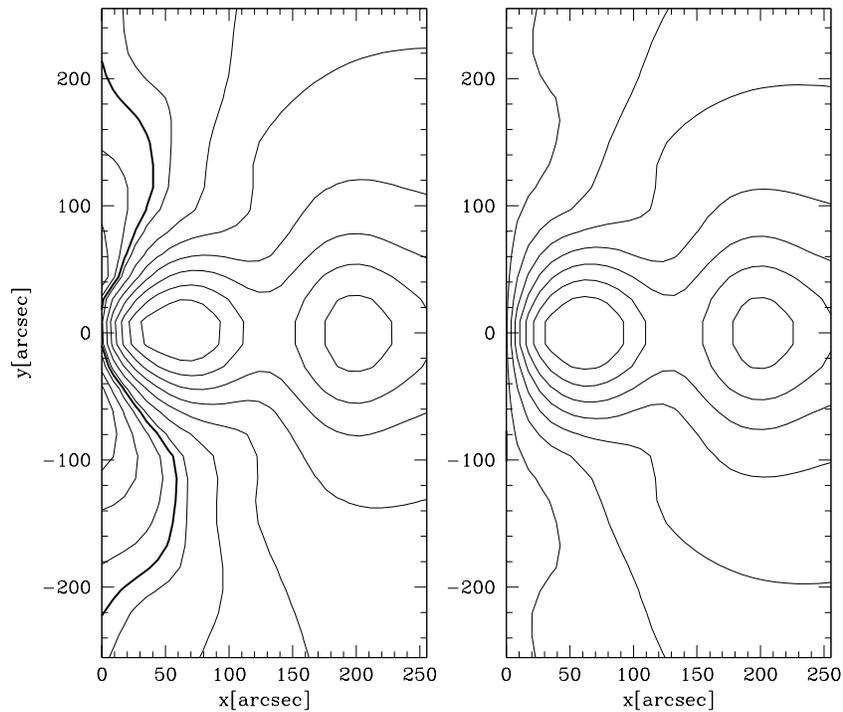}\caption{
Velocity field derived from the absorption line data combined with the
PNe data symmetrized around both major and minor axes.  The thin contours are 
spaced by $16.6~\kms$, starting at central $9.5~\kms$; thick contours
are at zero. The left panel
shows the velocity field including the minor axis absorption line
velocities; the right panel without.  Both have been constructed with
the smoothing algorithm and $\Lambda=1.2$ (see Section 3).  In both
panels there are thus 86 velocity points fitted in the region shown,
but the fit was actually done between $-250''$ and $250''$ for all 172 points
together with the unaltered respective absorption line data. This
ensures that in the right panel the velocities on the minor axis would
be zero. \label{v_field}
}
\end{figure}

We have argued earlier that the minor axis velocities are influenced
by non--equilibrium effects. We have therefore constructed velocity
fields both with and without the minor axis data. The essential
features of the velocity field including the minor axis data (left
panel of Fig.~\ref{v_field}) and that without (right panel) are
closely similar. A velocity peak at $60''$ corresponds to the maximum
value of the rotation curve reached in the integrated light
measurements, see Fig.~\ref{f3absv}; a second peak at $\sim 200''$
corresponds to the PNe rotation of $140\,\kms$ found at $4'$ in
Section 3. The dip between the two peaks is approximately $35\,\kms$
below the peaks while we estimate that the error on the velocities of
these major features is about $60\,\kms$.  Thus the dip is hardly
significant.  In gross terms, the velocity field in Fig.~\ref{v_field}
is similar to the velocity field plotted in Fig.~5 of Dehnen \&
Gerhard (1994) for a rotating isotropic oblate $\gamma=3/2$
model. This velocity field does not have rotation along the minor
axis.

\begin{figure}\plotone{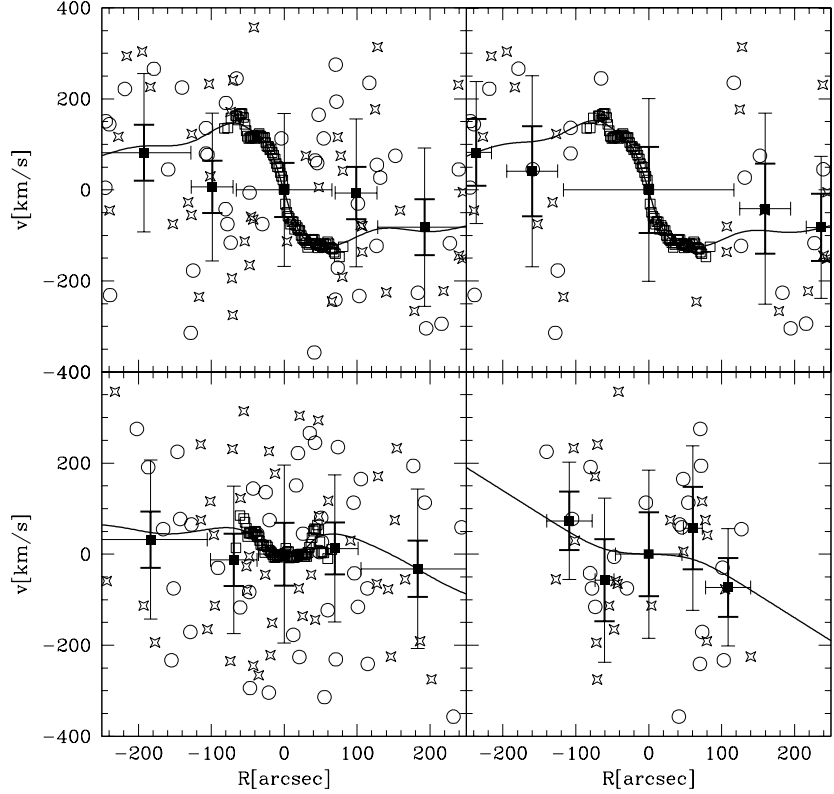}\caption{
Smoothed rotation curves (full lines) obtained from absorption line
data and PN velocities, with $\Lambda = 4$.  Top left: along
${\rm P.A.}=50^\circ$, the photometric major axis.  The entire point symmetric
data set of 86 PN velocities was used, with different symbols to
identify the actually observed 43 points (open circles), and their
symmetric counterpart (stars).  Bottom left: same, along the
photometric minor axis.  Top right: along the major axis, but only
absorption line velocities and PNe data points at heights less than
$73''$ above the major axis.  Bottom right: same, but only for PNe
data points at heights larger than $73''$ above the photometric major
axis. There are no absorption line data at these heights.  The squares
and thick error bars in each panel indicate the mean velocity and its
error, the thin ``error bars'' the calculated velocity dispersion
$\sigma$, all computed in a small number of bins each containing about
one fifth of the PNe.  The $1\sigma$ error on the mean velocity is
given by the velocity dispersion divided by $\sqrt{N_{\rm bin}}$,
i.e., about $60\, \kms$. \label{rot_new}
}
\end{figure}

Fig.~\ref{rot_new} shows the combined data projected onto the
photometric major axis at ${\rm P.A.} = 50^\circ$ and minor axis at ${\rm P.A.}
=140^\circ$ in much the same way as Fig.~\ref{f8rotcurve} for the PNe
alone.  Because the photometric major axis is misaligned with respect
to the LOMG for the PNe sample, the PN rotation velocities at several
arcmin radius in Fig.~\ref{rot_new} are noticeably smaller than those in
Fig.~\ref{f8rotcurve}. The difference is about $1\sigma$, similar to
the significance of the misalignment. The fitted mean rotation
curves are clearly dominated by the absorption line data in the region
where these are available. Especially from the top right panel
of Fig.~\ref{rot_new}, which plots the absorption line data and the lower
half of the PNe data within $73''$ of the major axis, it appears that
the PNe are not drawn from the absorption line velocity curve.
Probably this is because even the lower half of the PNe distribution
samples a different part of the velocity field than does the
major axis absorption line data. In other words, the gradients
in $z$ must be relatively strong.

It is not possible to recover the dependence of the velocity
dispersion on radius nonparametrically with a small sample like ours.
To check whether the velocity dispersion changes with radius at all,
we divide the PN data into two bins in major axis radius and determine
the \rms\ velocity dispersion with respect to the rotation curve
shown in Fig.~\ref{rot_new}. We calculate the variance of
the residual velocities both for the full PNe sample and for the lower
half of the sample close to the major axis.

\begin{figure}\plotone{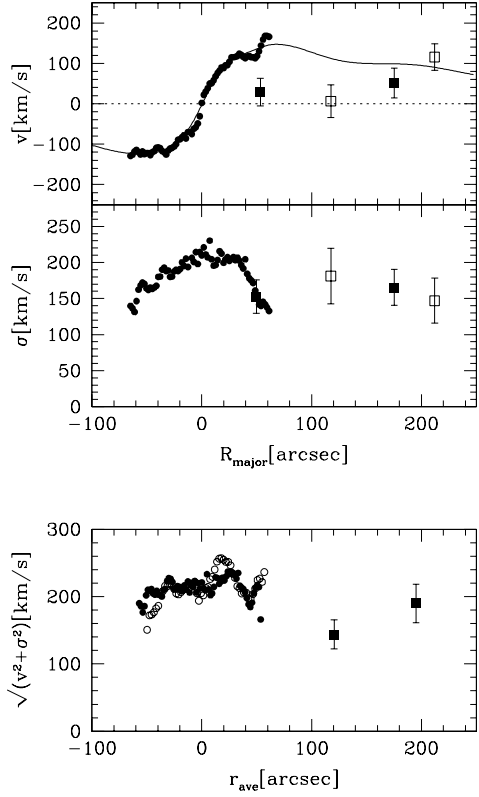}\caption{
Mean rotation velocity and velocity dispersion as a function of
distance, from absorption and PNe data.
Top panel: rotation velocity as a function of major
axis radius $R$. Curve is taken from Fig.~\ref{rot_new}, top left,
the two outer points are computed from the PNe velocities ordered
in $R$ and binned in equal halves. Middle panel: velocity
dispersion as function of $R$. The outer points with larger error
bars show the dispersion around the mean rotation velocity of
the PN sample in two bins (standard deviation). Filled squares
in both panels are obtained from Fig.~\ref{rot_new}, top left,
open squares from Fig.~\ref{rot_new}, top right (half of the
sample closer to the major axis). Bottom panel: mean square
absorption line velocities $\sqrt{v^2+\sigma^2}$ on the major
axis, rescaled to equivalent spherical radius by $r=\sqrt{0.7}R$
(filled circles). Same for minor axis absorption line velocities,
rescaled to equivalent spherical radius by $r=R/\sqrt{0.7}$
(open circles). Full squares show equivalent spherical PNe velocity
dispersions $(\sum (v_i-v_{\rm sys})^2/(N_{\rm bin}-1))^{1/2}$ in two bins.
Bin radii are mean values of $\sqrt{q(x_i^2+y_i^2/q^2)}$ for the
respective halves of the sample.\label{sigmafigure}
}
\end{figure}

Fig.~\ref{sigmafigure} shows the mean rotation velocity and velocity
dispersion along the major axis from the absorption line data and from
the binned PNe data. The lower panel of Fig.~\ref{sigmafigure} shows a
spherically averaged velocity curve, giving major axis and minor axis 
$\sqrt{v^2+\sigma^2}$ for the absorption line data, and binned values of 
$\sum_i (v_{i}-v_{\rm sys})^2$ for the PNe. 

The total variance with respect to the fitted rotation
velocity includes the measurement error $\sigma_{\rm obs}=42\,\kms$ and the intrinsic velocity dispersion $\sigma_{\rm int}$
in the galaxy, where
$\sigma^2 = \sigma_{\rm obs}^2 + \sigma_{\rm int}^2$.  For a variance of
$\sigma=165\pm 17\,\kms$ this results in an intrinsic velocity
dispersion of $\sigma_{\rm int}=159 \pm 17\, \kms$.  The error $\delta
\sigma_{\rm int}$ is calculated as the variance of the
$\chi^2$--distribution and is given by $\delta \sigma_{\rm int}=
\sigma_{\rm int}/\sqrt{2 (N-1)}$ with $N$ the size of the sample and
$N-1$ the number of degrees of freedom. The PNe dispersions plotted
in Fig.~\ref{sigmafigure} are the galaxy--intrinsic velocity
dispersions $\sigma_{\rm int}$.

Within the error bars, the major axis rotation
velocity outside $50''$ is constant at $v=110\,\kms$, the major axis
dispersion outside $50''$ (the central peak) is constant at $150\,\kms$,
and the spherically averaged mean squared velocity outside $40''$ is
constant at $v=200\,\kms$.

\newpage

\newpage

\newpage

\newpage

\section{Dynamics and Mass Estimates}

From the measured kinematics we can now estimate the dynamical
support by rotation. We do this in two ways. First, we evaluate
$$
 \left(v\over\sigma\right)^\ast \equiv {v_{\rm max}\over \=\sigma}
	/ \sqrt{\epsilon\over 1-\epsilon}	
$$
(Kormendy 1982). Here $v_{\rm max}$ and $\=\sigma$ are the maximum
observed rotation velocity and the luminosity--weighted mean of the
major axis dispersion inside $0.5 a_e$, and $\epsilon$ the
observed ellipticity. From Fig.~\ref{f3absv} we estimate
$v_{\rm max}=145\kms$ and $\=\sigma=185\kms$, whereas the mean
ellipticity within $a_e/2$ is $\epsilon\simeq 0.37$ (Schweizer 1980,
Caon \etal 1994). We thus find $(v/\sigma)^\ast=1.02$. This is somewhat
smaller than the corresponding value obtained for isotropic rotator models
($(v/\sigma)^\ast\gta 1.2$; Dehnen \& Gerhard 1994), but shows
that rotation is dynamically important.

Secondly, we use mass--weighted velocities from the kinematic data in
Fig.~\ref{sigmafigure} outside $50''$. For the mass--weighted projected
rotation velocity we take $\tilde v=110\kms$, and for the
projected dispersion $\tilde\sigma=150\kms$, giving $\tilde v /
\tilde\sigma = 0.73$.  The predicted value for an oblate isotropic
spheroid seen edge--on, using a projection factor of $\pi/4$ derived
by Binney (1978), is evaluated from equation (5.8) of Gerhard (1994):
$$
 {\tilde v\over \tilde\sigma} 
	\simeq {\pi\over 4} \sqrt{2 \left[(1-\eps)^{-0.9}-1\right] }.
$$
With an average ellipticity of $\epsilon=0.3$ in the range of
radii considered (Caon \etal 1994), this gives $\tilde v /
\tilde\sigma = 0.68$, i.e., $(\tilde v/\tilde\sigma )^\ast=1.07$.
Thus the main body of NGC 1316 is consistent with an
isotropic rotator with $\epsilon=0.3$ (inclination effects on
this conclusion are weak, Binney 1978).

We will now estimate the mass of NGC 1316 assuming that the galaxy 
is near--edge on and isotropic. The Jeans equation in the equatorial
plane $z=0$ of an axisymmetric system in cylindrical coordinates reads
\begin{equation}
 \left( \pbyd{\Phi}{R} \right)_{z=0} = {{\bar v_\phi}^2 \over R}
	-{\sigma_R^2 \over R} \left[  
	        \pbyd{\ln\rho}{\ln R} + \pbyd{\ln \sigma_R^2}{\ln R}
	      + 1-{\sigma_\phi^2 \over \sigma_R^2}  
	      + {R\over \sigma_R^2} \pbyd{\overline{v_R v_z}}{z}
							\right]_{z=0}.
\label{eqmass}
\end{equation}
Here ${\bar v_\phi}$ is the intrinsic rotational streaming velocity,
$\sigma_R$ and $\sigma_\phi$ are the radial and azimuthal velocity
dispersions, $\rho$ is the density, all referring to an equilibrium
tracer population that need not be self-consistent with the potential,
and $\Phi$ is the total gravitational potential. The PNe are not an
equilibrium distribution even if they emerge from an equilibrium
distribution of stars because of the selection effects; therefore they
are only used as tracers of the underlying velocity field.
The last derivative term depends on the orientation of the stellar 
velocity ellipsoid. This term vanishes for a flattened system with 
velocities isotropic in the meridional plane. The equatorial potential 
gradient may be associated with a `mass' 
$M_{\rm eq}(R)=R^2 (\pbyd{\Phi}{R})_{z=0}/G$, which
overestimates the real mass within $R$ by a factor of order unity
except in the case where the potential is spherical. For a spheroidal
density distribution with flattening $\epsilon=0.3$ and a density
profile inversely proportional to the elliptical radius squared, this
factor is $f_{0.3} = 1.13$.

To apply this equation, we again set the projected rotation velocity
$v = (\pi/4) v_\phi$. Isotropy implies $\sigma=\sigma_R=\sigma_\phi$,
and since the velocity dispersion profile is approximately flat,
only the first term in the square bracket survives. At a radius of
$200''=16\,\kpc=1.5 a_e$ on the major axis the logarithmic slope of the
density profile for a de Vaucouleurs model is $\simeq -2.9$, so the
equatorial mass is $ f_{0.3}^{-1} \times M_{\rm eq}(200'') =
2.9\times 10^{11}\msun$.

A similar estimate can be obtained from the spherical Jeans equation
$$
 M(r) = -{r\sigma_r^2\over G} \left( \tbyd{\ln\rho}{\ln r}
		+ \tbyd{\ln \sigma_r^2}{\ln r} + 2\beta \right)
$$
and the results in the bottom panels of Fig.~\ref{sigmafigure}.  Here
$\sigma_r(r)$ is the radial velocity dispersion, and $\beta(r) \equiv
1 - \sigma_\phi^2 / \sigma_r^2 < 1$ the anisotropy parameter. Assuming
again isotropy we have $\beta=0$, and, since the outer rotation
velocity field is not cylindrical, $\sigma = 150 \kms \lta \sigma_r
\lta \sqrt{v^2+\sigma^2} = 200\kms$, approximately independent of $r$.
Then we find $ 2.4\times 10^{11}\msun \lta M(200'') \lta 4.3\times
10^{11}\msun$.  A modest radial anisotropy in the outer regions of NGC
1316, such as inferred from modelling the line profiles in NGC 2434
(Rix \etal 1997), NGC 6703 (Gerhard \etal 1998) and NGC 1600 (Matthias
\& Gerhard 1998) would increase both estimates somewhat.

We do not have our own B-band photometry for \n1316, so we will use
published photometry for estimating the M/L ratio inside a radius $R
=200''$. If we compute the integrated luminosity inside $200''$ using
the aperture photometry listed in Lauberts \& Sadler (1984), we obtain $ L_B
= 3.7 \times 10^{10} L_\odot$\footnote{Integrated luminosities are computed 
within 3-D radius r, assuming an $r^{1/4}$ profile.} 
which implies a M/L$_B$ ratio of $7.7$.
To test for a gradient in the mass--to--light ratio, we similarly
compute integrated luminosities and masses interior to radii $45'',\,
90'',\, 143''$.  Masses are computed again from eq.~\ref{eqmass} with the
same assumptions as before; these are only applicable outside $\sim
50''$. We obtain masses $ 4.9 \times 10^{10}\msun$, $ 1.1 \times
10^{11}\msun$, $ 1.8 \times 10^{11}\msun$ inside these radii, and
cumulative mass--to--light ratios of 4.3, 5, 6.4, respectively,
compared to a value of 7.7 at $200''$.  

\begin{figure}\plotone{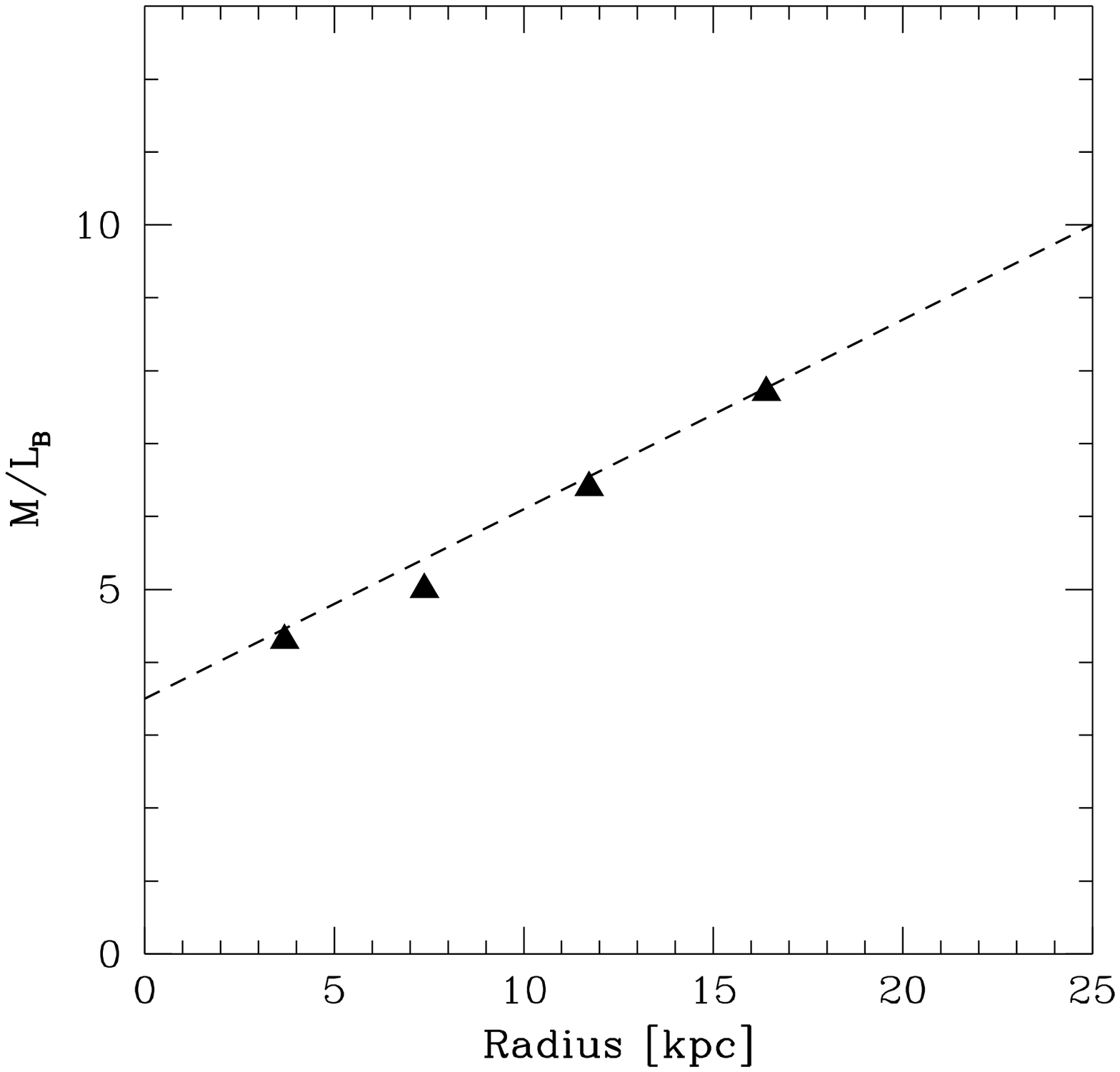}\caption{
The dashed line indicates the M/L$_B$ ratio as function of radius for
NGC~5128; triangles indicates the M/L$_B$ ratios obtained for
NGC~1316 at $R = 45'',\, 90'',\,143'',\, 200''$. \label{mlplot}
}
\end{figure}

NGC 1316 and~Cen A have somewhat similar morphological properties:
Figure~\ref{mlplot} shows that their radial distributions of M/L$_B$, as 
estimated from PN velocities in the same range of radius, 
are strikingly similar (see Hui \etal\ 1995 for the Cen~A data). We note that
the adopted distances for these two galaxies are on the same system
(planetary nebulae luminosity function and surface brightness fluctuations). 
The apparent radial increase of M/L in these two systems follows from (i) the 
observed $r^{1/4}$ or Hernquist distributions of surface brightness, (ii) the
observed approximate constancy of $\sigma^2+v^2$ with radius, and 
(iii) the assumption of isotropy at all radii: this assumption may
not be correct, so the precise shape of the M/L$(r)$ relation is still
somewhat uncertain. Nevertheless, the similarity of the M/L scale
for these two galaxies is remarkable.

\newpage

\section{Conclusions}

The \PN\ radial velocities show the presence of rotation in the outer parts
of \n1316, in the direction of an elongated structure visible in
Schweizer's deep image of this galaxy. This elongated structure is
orthogonal to the inner minor axis dust lane. The mean rotation
velocity along the galaxy's photometric major axis is constant outside
$50''$ at $v = 110 \kms$, out to the maximum observed radius of $200''=
1.8 R_e = 16$ \kpc.  Based on combined integrated light data in the
central regions plus PNe radial velocity data in the galaxy's halo the
velocity dispersion profile along the galaxy major axis is consistent
with being flat at $150 \kms$ between 4 \kpc\ and 16 \kpc.

From the Jeans equations and the assumption of isotropy in the
meridional plane, we estimate the mass of Fornax~A at several
radii. Within $200''$ the total mass is $2.9 \times 10^{11} M_\odot$,
and the corresponding integrated M/L ratio is $\simeq 8$.
The integrated M/L ratio decreases inwards, to about $4$ within
$50''$.  This apparent M/L gradient, which depends on the assumption 
that the system is isotropic at all radii, is like that of the other 
nearby large merger remnant Cen~A: both galaxies show a similar 
near-constancy of $\sigma^2+v^2$ with radius. We note that the inner 
M/L for these two galaxies is relatively low, about 3 to 4, which may
indicate that their stellar population is, in the mean, somewhat younger 
than usual.

With the extended kinematical data, one can measure the specific
angular momentum of \n1316. If one adopts the formulae given by Fall
(1983), the logarithmic values for the total luminous mass and the
specific angular momentum of \n1316\ are $\log ({\rm M/M}_\odot) = 
11.7$ \footnote{We assume M/L$_B = 6$ for early-type galaxies as in 
Fall (1983) to compute \n1316\ total luminous mass.} and 
$\log({\rm (J/M)}/\kms {\rm kpc}) = 3.5$ respectively. 
From  Fall (1983) diagram, spirals with total luminous mass as \n1316\ have 
$\log({\rm J/M})$ values in the range from 3.6 to 4.17, while spirals with 
total luminosity equal to the \n1316\ one have $\log({\rm J/M})$ values 
within the range from 3.38 to 4.05.
Therefore the kinematics of the outer halo of \n1316\ indicates that this 
early-type galaxy contains as much angular momentum as a giant spiral of 
similar luminosity.

The main result of this work is that even with 43 \PN radial
velocities we can extract interesting constraints on the dynamics of
the outer regions of giant early-type galaxies.  Given the limited
radial velocity sample that we were able to obtain for \n1316, this
work represent a case study.  When the 8 m telescopes provide us with
samples of discrete radial velocities larger by an order of magnitude,
the non-parametric approach combined with the use of integrated-light
absorption line data at small radii should allow a robust analysis of the 
system dynamics and its mass distribution out to several $R_e$.

\acknowledgments
The authors are grateful to the NTT team, the ESO staff at La Silla
and the MSSSO staff at Siding Spring for their help and support during
the observations.  They would like to thank F. Schweizer for
permission to reproduce the photographic print of NGC~1316, and the
referee, D. Merritt, for helpful comments on the non-parametric
method. M.A. and K.C.F. acknowledge support from the Australian
Government Department of Industry, Science and Tourism.  O.G. and
M.M. acknowledge support from the Swiss National Foundation under
grant 20-43218.95.  R.H.M. has been supported by the Deutsche
Forschungsgemeinschaft through Grant SFB (Sonderforschungsbereich)
375. M.A. would like to acknowledge G. Busarello and G. Longo for the
use of the adaptive filtering algorithm.



\end{document}